**Title:** What confines the rings of Saturn?

**Authors:** Radwan Tajeddine[1,*], Philip D. Nicholson[2], Pierre-Yves Longaretti[3,4], Maryame El Moutamid[1], Joseph A. Burns[2,5]

*Corresponding Author: Tajeddine@astro.cornell.edu

**Affiliation:**

[1]Center for Astrophysics and Planetary Science, Cornell University, Ithaca, NY 14853, USA

[2]Department of Astronomy, Cornell University, Ithaca, NY 14853, USA

[3]Université Grenoble Alpes (UGA), France

[4]CNRS/INSU – Institut de Planétologie et d'Astrophysique de Grenoble (IPAG), UMR 5274, France

[5]College of Engineering, Cornell University, Ithaca, NY 14853, USA

**Abstract**

The viscous spreading of planetary rings is believed to be counteracted by satellite torques, either through an individual resonance or through overlapping resonances. For the A ring of Saturn, it has been commonly believed that the satellite Janus alone can prevent the ring from spreading via its 7:6 Lindblad resonance. We discuss this common misconception and show that, in reality, the A ring is confined by the contributions from the group of satellites Pan, Atlas, Prometheus, Pandora, Janus, Epimetheus, and Mimas, whose cumulative torques from various resonances gradually decrease the angular momentum flux transported outward through the ring via density and bending waves. We further argue that this decrease in angular momentum flux occurs through 'flux reversal'.

Furthermore, we use the magnitude of the satellites' resonance torques to estimate the effective viscosity profile across the A ring, showing that it decreases with radius from ~50 cm$^2$ s$^{-1}$ to less than ~10 cm$^2$ s$^{-1}$. The gradual estimated decrease of the angular momentum flux and effective viscosity are roughly consistent with results obtained by balancing the shepherding torques from Pan and Daphnis with the viscous torque at the edges of the Encke and Keeler gaps, as well as the edge of the A ring.

On the other hand, the Mimas 2:1 Lindblad resonance alone seems to be capable of confining the edge of the B ring, and contrary to the situation in the A ring, we show that the effective viscosity across the B ring is relatively constant at ~24-30 cm$^2$ s$^{-1}$.

# 1. Introduction

   Like protoplanetary and accretion disks, planetary rings are subject to radial spreading. Such a phenomenon happens whenever inner ring particles exchange angular momentum with outer ones. Since the inner ring particles orbit the planet (or star) faster then the outer ones, the slower-moving outer particles gain angular momentum causing them to drift outward, while the faster-orbiting inner particles lose angular momentum and drift inward. This angular momentum exchange is accompanied by a net loss of kinetic energy and translates on the largest scale into radial spreading of the ring in both directions, toward and away from the planet. For a dense ring, such diffusive spreading can be characterized by the ring's effective viscosity. For the A ring of Saturn, the effective viscosity is expected to be dominated by self-gravitational transport with an expected magnitude of the order of $\nu \sim 100 - 200$ cm$^2$ s$^{-1}$ (Daisaka et al, 2001; Yasui et al, 2012). With a radial width $\Delta r = 15,000$ km, the spreading timescale is $(\Delta r)^2/\nu \sim 7 \times 10^8$ years. This represents the timescale for the ring to spread from a hypothetical narrow band at a radius of ~130,000 km to its current dimensions. A more sophisticated modeling of the viscous evolution of the rings of Saturn (Salmon et al. 2010) suggest a spreading timescale of ~$10^8$ years. If the system did not have any satellites these numbers would have represented an upper limit on the age of the A ring. However, the satellites of Saturn are believed to play an important role in stopping the rings from spreading, or significantly reducing the rate at which they do so, and therefore extending the age of the rings.

   The confinement of the rings of Saturn was for a long time an unsolved matter. The satellite Atlas might be expected to play a major role in confining the A ring through the shepherding mechanism since it is the closest satellite to its outer edge. However, Voyager imaging and occultation data showed that the edge of the A ring appears to be in a 7:6 Inner Lindblad Resonance (ILR) with the more massive satellite Janus (Porco et al. 1984). Thus, it has been generally believed that the torque exerted by this resonance confines the entire A ring. However, this interpretation has recently been questioned because of the large radius uncertainties in the Voyager imaging data and the fact that the 7:6 resonance was located well outside the A ring edge at the period of the Voyager flybys. Indeed, Janus moves on a horseshoe orbit, switching orbits every four years with the smaller satellite Epimetheus. As a consequence, the 7:6 resonance moves back and forth relative to the edge of the A ring. El Moutamid et al. (2016) showed that the 7-lobed resonantly-forced pattern disappears when the resonance location is far from the edge, confirming the finding by Spitale & Porco (2009) that the edge of the A ring is in resonance with Janus only during the half of the eight-year libration period when this satellite is closer to the planet.

   On the other hand, the B ring's outer edge has been clearly shown to be controlled by the 2:1 ILR with Mimas (Porco et al. 1984; Spitale & Porco, 2010; Nicholson et al. 2014). This resonance, the strongest anywhere in Saturn's rings, is believed to prevent the B ring from spreading outwards and also to be responsible indirectly for the existence of the Cassini division (Goldreich & Tremaine, 1978a; see also Hahn et al, 2009).

   As we discuss in more detail below, ring confinement depends mainly (among other parameters) on the confining satellite's mass and the effective viscosity of the ring.

Thanks to the Cassini mission, the masses of the satellites of Saturn are now well known (Jacobson et al. 2006, 2008; Weiss et al. 2009). However, the viscosity of the rings of Saturn is a parameter that remains poorly understood. In the A ring, estimates from Voyager data of the viscosity (inferred mainly from the damping of density waves in occultation profiles) range from ~65 cm$^2$ s$^{-1}$ to more than 2000 cm$^2$ s$^{-1}$ (Esposito et al. 1983, Lissauer et al. 1984, Shu et al. 1985, Chakrabarti 1989). Recently, Tiscareno et al. (2007) applied the same method to weaker waves seen in Cassini ISS images to suggest that the viscosity in the A ring increases gradually from ~25 cm$^2$ s$^{-1}$ near the inner edge of the ring to ~300 cm$^2$ s$^{-1}$ near the inner edge of the Encke gap. Although Cassini did observe density waves in the trans-Encke region of the A ring, inferring the ring viscosity from wave damping in this region is complicated because of wave interference and non-linearity issues. However, these estimates of the viscosity are only representative of the viscosity in perturbed regions as we discuss in Section 5.

On the other hand, Spitale & Porco (2010) measured the phase lag between the longitude of the *m*=2 pattern at the outer edge of the B ring and that of Mimas to obtain the only estimate of the viscosity in B ring of ~20 cm$^2$ s$^{-1}$ (applying the relation between the phase lag and the viscosity given by Borderies et al. 1982, which was intended only as a crude approximation). This method is based on angular momentum transfer between the ring edge and Mimas, and as we also discuss in Section 5, the value and even the definition of viscosity depend strongly on the particular process being studied.

In this work, we study the effect of resonance torques due to multiple satellites on the confinement of the A and B rings of Saturn, as well as on the orbital evolution of the satellites. In Section 2 we summarize the different torques exerted by satellites on the rings and vice versa, and introduce the main problem of confinement and angular momentum transport across the rings; we also derive an expression for the angular momentum transfer via bending waves. In Section 3, we first study the variation of the angular momentum flux (henceforth abbreviated as AMF) and the implied effective viscosity of the A ring as one crosses the many first and second order Lindblad and vertical resonances in this region. We then carry out a similar calculation for the B ring.

In Section 4 we discuss the response of the rings to satellite resonant torques and discuss their effects on the orbital evolution of the satellites. Some implications of our results are discussed in Section 5 and our conclusions are summarized in Section 6.

## 2. Theoretical background: Satellite torques

Voyager and Cassini observations have revealed many gaps and ringlets with sharp edges in the rings of Saturn (see the review by Nicholson et al. 2017). The maintenance of some of these sharp edges is related to satellites; for example, Pan and Daphnis are located at the centers of the Encke and Keeler gaps, respectively, while Janus and Mimas are thought to confine the A and B ring outer edges through discrete ILRs, as discussed above. However, the origin of many other sharp edges (e.g., several gaps in the Cassini Division and the C ring) in Saturn's rings remains unclear.

Satellites can control and sharpen ring edges in one of two ways: either through an individual resonance, like the Janus 7:6 and the Mimas 2:1 resonances, or via the overlap of many, closely-spaced resonances (generally referred to as "shepherding"), such as the situation with Pan and Daphnis. Much of the theoretical background for this picture was developed by Borderies, Goldreich and Tremaine (1982, 1989); for a recent summary and in-depth discussion of the uncertainties and limitations of this model, the reader is referred to the review by Longaretti (2017a).

## 2.1 Lindblad resonances

When a ring particle is in an Inner/Outer Lindblad Resonance (ILR/OLR) with a satellite, its mean motion obeys the relation

$$(m-1)n + \dot{\varpi} - (m+k)n_s + k\dot{\varpi}_s = 0, \tag{1}$$

where $m$ is the azimuthal wave number ($m>0$ for an ILR and $m<0$ for an OLR) and $k$ represents the resonance order ($k=0$ for first order, $|k|=1$ for second order, etc.). As usual, $n$ and $n_s$ are the mean motions of the ring particle and satellite, respectively, and $\dot{\varpi}$ and $\dot{\varpi}_s$ are the local and the satellite's apsidal precession rates due to Saturn's gravity field, respectively (For third and higher-order resonances, $\dot{\varpi}_s$ may be replaced by $\dot{\Omega}_s$, the satellite's nodal regression rate, subject to the restriction that the coefficient of $\dot{\Omega}_s$ be even). Expressions for these angular frequencies are given by French et al. (1982), Nicholson & Porco (1988), and Borderies-Rappaport & Longaretti (1994):

(2)
$$n \approx \sqrt{\frac{GM_p}{a^3}}\left\{1 + \frac{3}{4}J_2\left(\frac{R}{a}\right)^2\left(1+4e^2-16\sin^2 i\right) - \left[\frac{15}{16}J_4 + \frac{9}{32}J_2^2\right]\left(\frac{R}{a}\right)^4 + \left[\frac{27}{128}J_2^3 + \frac{45}{64}J_2 J_4 + \frac{35}{32}J_6\right]\left(\frac{R}{a}\right)^6\right\}$$

$$\dot{\varpi} \approx \sqrt{\frac{GM_p}{a^3}}\left\{\frac{3}{2}J_2\left(\frac{R}{a}\right)^2\left(1+2e^2-2\sin^2 i\right) - \frac{15}{4}J_4\left(\frac{R}{a}\right)^4 + \left[\frac{27}{64}J_2^3 - \frac{45}{32}J_2 J_4 + \frac{105}{16}J_6\right]\left(\frac{R}{a}\right)^6\right\}.$$

(3)

where $M_p$, $R$, and $J_n$ are the mass, equatorial radius and the zonal gravity harmonics of Saturn, (Jacobson et al., 2006) and $i$ and $e$ are the inclination and eccentricity of the ring particle (or satellite) orbit, respectively, and $G$ is the gravitational constant. The original formulae given by Nicholson & Porco (1988) contain additional terms that describe perturbations by external satellites, but these are negligible for ring particles and have been ignored for the purpose of this study. In these expressions, the quantity $a$ is the local mean (or epicyclic) radius (distance from the planet), as defined by Borderies-Rappaport and Longaretti (1994), rather than the osculating semi-major axis used in most celestial mechanics texts. For a ring particle, the terms proportional to $e^2$ and $\sin^2 i$ can also be neglected.

For a first-order ILR (OLR), after $m$ orbits the ring particle will return to the same longitude relative to the satellite, at the same mean anomaly, while the latter will have orbited Saturn $m$-1 ($m$+1) times. A similar, but somewhat more complicated, situation applies to higher-order resonances for which the mean anomaly of the satellite is also important. In all cases, the azimuthal parameter $m$ differs by 1 from the coefficient of $n$ in Eq. (1). The order of the resonance is the difference between the coefficients of $n$ and $n_s$, higher-order resonances generally being weaker than first-order ones.

At a Lindblad resonance, perturbations from the satellite excite the ring particles' eccentricities. As a result, in the reference frame of the perturbing satellite, the ring section locked in resonance develops an $m$-lobed pattern, representing the $m$ apoapses and periapses that each ring particle goes through before returning to the same longitude it started from in the satellite's frame[1]. As the ring particles oscillate radially, their perturbations are transferred via self-gravity to the neighboring ring particles closer to the satellite and farther from the resonance location. Because these neighboring ring particles orbit the planet at a slower (faster) rate than those locked in the ILR (OLR), a trailing spiral density wave is created, which propagates radially outward (inward) across the ring in the direction of the perturbing satellite. The wave itself carries angular momentum: negative for an ILR and positive for an OLR (Shu 1984; Toomre 1969; Dewar 1972; Goldreich and Tremaine 1978b).

Subsequently, the density wave is damped by collisions between the ring particles as it propagates through the ring, causing these particles to lose (gain) angular momentum and drift back towards the resonant location. The net result is a transfer of angular momentum from the satellite to the ring when the density wave is damped. On other hand, a corresponding reaction torque is exerted by the ring on the satellite. At an ILR, the ring loses angular momentum while at an OLR it gains it. The rate of transfer of angular momentum by a single resonance is given by the expression (see Appendix A for details),

$$T_R = \mp \frac{4\pi^2}{3} \frac{m}{m-1} \Sigma \left( a^2 \beta n \frac{M_s}{M_p} e_s^{|k|} A_{m,k} \right)^2 \qquad (4)$$

where the minus (plus) sign applies to an ILR (OLR). Here $\beta = a/a_s$, and $a_s$, $M_s$ and $e_s$ are the semi major axis, mass and eccentricity of the perturbing satellite, respectively; $\Sigma$ is the mean surface mass density of the ring at the resonance location. The expression for the dimensionless quantity $A_{m,k}(\beta)$ is given in Appendix A; it depends on the Laplace coefficients $b_\gamma^m(\beta)$ and their derivatives that can be calculated numerically (Brouwer & Clemence 1961) using the definitions given in Murray & Dermott (1999, Eq. 6.67). In the limit $m >> 1$, $m/(m-1) \approx 1$, $\beta \to 1$ and $A_{m,0} \approx 0.8m$ for a first order resonance ($k$=0), and $A_{m,1} \approx 0.48m^2$ for a second order resonance ($k$=1). Eq. (4) then becomes

$$T_R = \mp f e_s^{2|k|} \left( \frac{M_s}{M_p} \right)^2 \Sigma n^2 a^4, \qquad (5)$$

---

[1] This description strictly applies only to first-order resonances; for higher-order Lindblad resonances the satellite is replaced by a perturbing potential component that rotates with a "pattern speed" $\Omega_p = [(m+k)n_s - k\dot{\varpi}_s]/m$.

where, $f \approx 8.6m^2$ for a first order resonance ($k = 0$) and $f \approx 3m^4$ for a second order resonance ($k = 1$). This expression is the same as the one derived by Goldreich and Tremaine (1978b, 1979), Shu et al. 1985, Meyer-Vernet and Sicardy (1987), and Longaretti (2017). For low $m$, Eq. (5) is accurate to within a factor of 2 – 3 and should be used only to get a rough estimate of the satellite torque.

Note that the resonant torque depends on the square of the satellite's mass, as is the case for the shepherding torque discussed below. The reason is the same as for tidal torques: the torque results from the gravitational interaction of the satellite with the perturbation that it raises in the rings.

The torque expressions above apply to a situation where the satellite resonance is located within the ring. However, when a resonance is located at a ring edge, the magnitude of the torque also depends on the phase lag $\Delta$ between the longitude of the apoapse/periapse of the $m$-lobed pattern and that of the perturbing satellite. A rough estimate of the torque can be obtained by multiplying the torque expression in Eq. (4) by $\sin(m\Delta)$. Borderies et al. (1982) derived a crude expression for the Mimas 2:1 ILR torque at the outer edge of the B ring; this is further discussed in section 3.3.

## 2.2 Viscous angular momentum transport

The AMF transported through the ring due to keplerian shear is given by the expression (Borderies et al. 1984, Lynden-Bell & Pringle 1974),

$$T_\nu = 3\pi\nu\Sigma na^2, \qquad (6)$$

where $\nu$ is the effective viscosity of the ring. More precisely, $\nu \equiv \nu_0 G(q)$, where $\nu_0$ is its unperturbed effective viscosity and $G(q)$ is a dimensionless factor depending on the effective local level of streamline perturbation $q$ (see sect. 5 for further details). If resonances do not overlap, $q=0$ and $G(q)=1$ in between resonances and this relation reduces to the usual one. If resonances overlap (for example if different density wave trains are present so that an unperturbed region cannot be found in some section of the ring), the effective $q$ is expected to be non zero and $G(q)$ has a more complex behavior. It may potentially reaching zero if AMF reversal takes place. Note that it is always possible to define $\nu_0$ and $\nu$ such that Eq. (6) holds. Such effective viscosities can be defined even if the dominant small-scale angular momentum transport mechanism is not collisional. The form of $G(q)$ depends on this mode of transport (see Longaretti 2017 for more details on this issue). The present work aims at characterizing $\nu$ and the potential relevance of AMF reversal to the present analysis will be further discussed in the discussion Section 5.2.

In the presence of an ILR, if the torque from the satellite resonance is stronger than the ring's viscous torque (|TR| ≥ Tν) the ring particles in the region perturbed by the density wave lose angular momentum faster than it can be replaced by the viscous flux and therefore move inward, creating a gap with a sharp inner edge. In this case, TR represents the maximum satellite torque that can be exerted on the ring, but Eqs. (4) and (5) can greatly overestimate the actual torque once a gap is formed (see section 3.3). At an OLR, if the satellite torque is again larger than the viscous torque, but positive, then the ring particles drift outwards, creating a gap with a sharp outer edge. In this manner a single satellite could, in principle, maintain a gap in a broad ring via an ILR at the gap's inner

edge and an OLR at its outer edge. In such a situation, the viscous flux of angular momentum is removed from the ring at the inner edge of the gap by the satellite resonance and then transferred back to the ring at the gap's outer edge. The net torque on the satellite is zero when the satellite reaches its equilibrium position.

In the case of a narrow ring bounded by a pair of satellites (such as the uranian ε ring), the inner edge of the ring is defined by an OLR with the interior satellite, which exerts a positive torque on the ring. This angular momentum is then carried across the ring by the viscous torque until it is removed from the ring at its outer edge by an ILR with the exterior satellite. In this case, the reaction torques on the satellites will cause them to slowly recede from the ring, with a net transfer of angular momentum from the inner satellite to the outer one. Detailed modeling suggests that flux reversal also plays a significant role here in reducing the effective viscosity of the perturbed ring (Borderies et al. 1984).

## 2.3 Overlapping resonances

Another mechanism of sharpening a ring edge, and of maintaining a narrow gap in a broad ring, is shepherding by a nearby satellite. From Eq. (1) we see that the resonance locations asymptotically approach the satellite as $m$ increases. For very high $m$, individual first-order resonances overlap in a small region close to the edge when the resonance spacing becomes smaller than the characteristic resonance width of $\sim(M_S/M_P)^{1/2}a$, i.e. when $\frac{M_P}{m^4 M_S} \sim< 1$ (Porco and Goldreich 1987, Borderies et al. 1989, Longaretti 2017). As a result, the satellite exerts a shepherding torque that is a sum over the torques associated with the overlapping resonances, leading to the expression (Lin and Papaloizou 1979; Goldreich & Tremaine, 1980, 1982),

$$T_S \approx \frac{g^2}{6}\left(\frac{a}{\Delta a}\right)^3 \left(\frac{M_s}{M_p}\right)^2 \Sigma n^2 a^4, \qquad (7)$$

where $\Delta a$ is the distance between the satellite and the ring edge and where $g = \frac{8}{9}[2K_0(2/3) + K_1(2/3)] \cong 2.24$. In practice, this distance adjusts itself until a balance between the satellite's and the ring's viscous torques is achieved.

However, there is a minimum satellite mass for which a complete gap is opened in a broad ring. If the satellite is smaller than this, then a partial gap forms but closes again after the encounter with the satellite, due to viscous spreading of the ring. Such is believed to be the origin of the propellers found in the A ring (Tiscareno et al. 2006; 2008; Sremčević et al. 2007). In the A ring, this minimum mass corresponds to a satellite of radius $R_{min} \sim 3.3(\nu/100 \text{ cm}^2\text{g}^{-1})^{1/3}$ km (Nicholson et al. 2017). For larger satellites such as Daphnis and Pan, the ring does not have enough time to close the gap between successive encounters with the satellite, resulting in the formation of the Keeler and Encke gaps, respectively.

## 2.4 Vertical resonances

A satellite can transfer negative angular momentum to the ring through another mechanism. A satellite with a non-negligible inclination excites the inclinations of the ring particles at an inner or outer vertical resonance (IVR/OVR). The resonant location is given by a slight variant of Eq. (1), with $\varpi$ replaced by $\dot{\Omega}$, but in this case the parameter $k$ must be an odd number (ie., +/-1, +/-3, …). In a self-gravitating ring, this results in a bending wave that propagates away from the resonance in a direction *away* from the perturbing satellite, opposite to that of density waves (Shu 1984). The best-known example of such a wave is that driven at the Mimas 5:3 IVR in Saturn's A ring, which was originally studied in Voyager images (Shu et al 1983). As in the case of a density wave driven at an ILR, the bending wave generated at an IVR also carries negative angular momentum, leading to a negative torque on the rings as it is damped. The signs are reversed at an OVR.

In Appendix A2 we derive an expression for the torque resulting from a second-order IVR/OVR (there is no first order torque as the satellite needs to be inclined in the first place):

$$T_{VR} = \mp \frac{\pi^2}{12} \frac{m}{m-1} \Sigma \left( a^2 \beta n \frac{M_s}{M_p} b_{3/2}^m I_s \right)^2 \tag{8}$$

where $I_s$ is the inclination of the satellite and $b_{3/2}^m(\beta)$ is a Laplace coefficient that must be calculated numerically (See Murray & Dermott 1999 Eq, 6.67). The minus (plus) sign applies to an IVR (OVR). In the limit $m \gg 1$, $m/(m-1) \approx 1$, $\beta \to 1$, and $b_{3/2}^m \approx 1.1 m^2$ Eq. (8) then becomes approximately

$$T_{VR} \approx \mp m^4 \Sigma a^4 n^2 \left( \frac{M_s}{M_p} \right)^2 I_s^2 \tag{9}$$

In comparison to the torque from a first order ILR/OLR in Eqns. (4) and (5), Eqs. (8) and (9) contain an additional factor $I_s^2$ that makes the torque from the strongest vertical resonances smaller than that from first order Lindblad resonances with the same satellite, but roughly equivalent to that of second order Lindlad resonances (which depend on $e_s^2$). For this reason, there are no known examples in the solar system where a satellite maintains a ring edge through a vertical resonance.

## 3. Angular momentum fluxes across the rings

We can apply the formulae in Section 2 to examine the ring confinement mechanisms at the outer A ring, Keeler gap, and Encke gap edges. In this section we also study the effect of other satellite resonances on the AMF and the viscosity of the A and B rings and examine the role they play in their confinement. Ultimately we are trying to answer the following questions: (1) what mechanism actually confines the rings of Saturn? and (2) what is the rings' effective viscosity and how does it vary with radius across the system?

To calculate the satellite torques in the A ring, we use a simplified surface mass density profile based on the results of Tiscareno et al. (2007). For radii between 124,000 km and 132,000 km, we assume that $\Sigma = 33.7 + 1.3 \, (a_{1000} - 124)$ g/cm$^2$, where $a_{1000}$ is the radius in thousands of kilometers. Tiscareno & Harris (2017) found that the surface mass density beyond this radius decreases gradually to reach $\Sigma = 15$ g/cm$^2$ at the outer edge of the A ring. Thus, we adopt a rough model of $\Sigma \approx 44.3 - 6.1$ ($a_{1000}$ - 132) g/cm$^2$ for radii beyond 132,000 km (see Tiscareno & Harris 2017 for more accurate values for the surface mass density across the A ring). We use Eq. (2) to calculate the mean motions of ring particles at different radii.

### 3.1 Viscosity of the outer A ring

We begin with the outermost part of the A ring, where we can apply Eqs. (4-7) to calculate the required ring viscosity that results in a balance between the satellite and the ring torques at the edge of the A ring and at the Keeler and Encke gaps. The outer edge of the A ring is located at a mean radius of 136,770 km and coincides, for one-half of the time, with the 7:6 ILR with Janus, as discussed in Section 1 above. Using Eqs. (4) and (6), with $m = 7$, we obtain an effective ring viscosity at the outer edge of the A ring of ~11 cm$^2$ s$^{-1}$. However, we note that Eq. (4) represents the maximum torque from a satellite Lindblad resonance, as derived from density wave theory, but that the actual torque exerted on the ring edge will be reduced with respect to the full linear torque due to the truncation of the ring at the edge. As described above the expression for such an edge torque depends on the phase lag between the ring $m$-lobed streamlines and the perturbing potential. Such a lag has not been measured for the A ring, but appears to be small (R.G. French and N. Cooper, priv. comm.). Because of this reduction in the actual torque with respect to the full linear torque, our estimated effective viscosity represents the *maximum* value and the actual value may be much lower than 11 cm$^2$ s$^{-1}$.

For Daphnis and the Keeler gap (radius ~136,505 km), as well as for Pan and the Encke gap (radius ~133,854 km), the overlapping resonance model is more appropriate than the discrete resonance one. Applying Eqs. (6) and (7) to these features, and using the estimates of mass from Weiss et al. (2009), we obtain an effective ring viscosity of ~14 cm$^2$ s$^{-1}$ and ~64 cm$^2$ s$^{-1}$, respectively. Note the decrease in the effective viscosity (and the angular momentum flux) as function of ring radius.

Our estimates of viscosity from torque balancing are much smaller than those estimated from the damping of density waves by Tiscareno et al. (2007), who obtained values of ~300 cm$^2$ s$^{-1}$ in the near-Encke region. In fact, if we use the latter value then none of the satellites mentioned above would be capable of maintaining the ring or gap edges, yet occultation data (reviewed by Nicholson et al. 2017) clearly show that all five edges of the outer A ring, Keeler gap and Encke gap are very sharp. Eqs. (4-7) depend on several physical parameters of Saturn, as well as of its satellites and rings, but only the two ring parameters, $\Sigma$ and $\nu$, are at all uncertain. The surface mass density appears in each of Eqs. (4-7) and thus cancels out when comparing the viscous torque in the rings to those from the satellites. This leaves the ring's effective viscosity as the only adjustable parameter in this model.

## 3.2 Collective resonant confinement of the A ring

As noted above, if the resonant torque from a satellite's ILR is stronger than the local viscous torque in the ring, we expect that a gap with a sharp inner edge should open at the resonance location. However, even if the resonant torque is smaller and no gap forms, then the satellite resonance will still exert a negative torque on the ring, via a density wave, which will reduce the outward flux of angular momentum through the rings. In this scenario, the Janus 7:6 ILR is responsible only for removing the flux that has 'survived' all of the previous similar satellite resonances. To quantify this picture, we calculate the variation of the AMF across the A ring taking into account *all* of the first and second order Lindblad resonances due to the satellites Pan, Atlas, Prometheus, Pandora, Janus, Epimetheus, and Mimas., as well as the strongest (i.e., second-order) vertical resonances due to Mimas.

By substituting Eqs. (2) and (3) into Eq. (1) and solving for *a*, we can calculate the resonance locations for each satellite, for different orders and *m* numbers. Table 1 lists the first and second order resonances, as well as the mass, eccentricity and semi-major axis for each satellite contributing to the viscous flux across the A and B rings (a full map of resonances with their strengths is given in Tiscareno & Harris 2017). We then use the expression in Eq. (4), along with the simple linear expressions for the surface mass density given above, to estimate the torque exerted on the ring by each resonance.

We do not consider the resonances of Daphnis, because its mass is so small that the associated torques are negligible. In addition, Atlas' and Pan's second order resonances were not included either because of their negligible torques (due to small masses and eccentricities), and for their first-order resonances the absolute *m* values are limited to a maximum of 100 (for higher values of *m* the Pan resonances start overlapping near the Encke gap and the angular momentum transfer is then described by the overlapping resonance torque due to a close satellite). Pan and Daphnis are unique in having both ILR (*m>0*) and OLR (*m<0*) resonances in the main rings because they are located within gaps in the A ring. For this reason, each serves as a bridge to transfer the AMF from one ring edge to the other via their overlapping resonance torques. These torques thus are equal and opposite on the inner and outer gap edges and need not be considered when studying the large-scale AMF through the A ring.

Similar to the situation for Janus, as Epimetheus moves in and out on its horseshoe orbit, so does its 7:6 ILR. It is outside the A ring when the Janus 7:6 ILR is at the outer edge of the ring, and then moves within the rings when the Janus 7:6 ILR moves out. Thus, only one of the Janus and Epimetheus 7:6 resonances affects the rings at any given time and their time-averaged torque is one-half of that given by Eq. (4). Note that the decay time of the 7-lobed perturbation is longer than the 4-year libration period of the coorbital satellites. Consequently, expressions of the torque derived in a stationary context do not formally apply. However, it is known that when the full *linear* torque applies (i.e., without a ring edge) the torque is independent of the specific physical process of transfer of the torque to the ring (Meyer-Vernet & Sicardy 1987). We may expect that the similar features hold in the present context and that the stationary edge torque estimate should apply half the time as mentioned above, although edge torques do explicitly depend on the ring internal dissipation. These points will require formal theoretical backing in the future.

| Satellite | $a_s$ (km) | $e_s$ ($10^{-3}$) | $M_s$ ($10^{15}$ kg) | A ring resonances | | B ring resonances | |
|---|---|---|---|---|---|---|---|
| | | | | First order | Second order | First order | Second order |
| Enceladus | 237,948 | 4.7 | 108022.0 | – | – | – | 3:1 |
| Mimas | 189,176 | 19.6 | 37493.0 | – | 5:3 | 2:1 | 4:2 |
| Epimetheus | 151,466 | 9.8 | 526.6 | 4:3 – 7:6 | 11:9 – 13:11 | 2:1, 3:2 | 4:2 – 6:4 |
| Janus | 151,412 | 6.8 | 1897.5 | 4:3 – 7:6 | 11:9 – 13:11 | 2:1, 3:2 | 4:2 – 6:4 |
| Pandora | 141,710 | 4.2 | 137.1 | 5:4 – 19:18 | 10:8 – 37:35 | 3:2, 4:3 | 5:3 – 8:6 |
| Prometheus | 139,380 | 2.2 | 159.5 | 6:5 – 35:34 | 11:9 – 69:67 | 3:2, 4:3 | 5:3 – 8:6 |
| Atlas | 137,670 | 1.2 | 6.6 | 6:5 – 100:99 | – | 3:2, 4:3 | – |
| Pan | 133,854 | 0.01 | 4.3 | 8:7 – 100:99, -99:-100 – -27:-28 | – | 3:2, 4:3 | – |

**Table 1.** Physical and orbital parameters of eight Saturnian satellites with resonances located in the A and B rings. The final columns show the ranges of first order and second order ILRs in each ring. For Prometheus, for example, the first order resonances in the A ring have $m$ numbers ranging from 6 to 35 while the second order ILRs of the same satellite range from $m$ = 10 to 68 as the radius increases. The masses of Enceladus, Mimas, Epimetheus, Janus, Prometheus, Pandora, and Atlas are taken from Jacobson et al. (2006; 2008), and the mass of Pan is from Weiss et al. (2009). The semi-major axes

of Janus and Epimetheus represent the periods in the configuration of their mutual horseshoe orbit when Janus is closer to the ring (see text).

In addition to the 5:3 ILR, Mimas has a relatively strong 5:3 IVR at a radius of ~131900 km (due to its inclination of ~1.6°). Therefore, we apply Eq. (8) to compute the torque exerted by this resonance.

Because we know that the AMF transmitted through the ring must reach zero after encountering the Janus 7:6 ILR at the outer edge of the A ring, we set this number as a starting point (see section 3.1) and estimate the viscous flux going backwards in radius across the A ring. The flux increases (with decreasing radius) by $T_R$ (Eq. 4) at each resonant location and by $T_{VR}$ at the Mimas 5:3 bending wave (Eq. 8). We use the same simplified piecewise-linear profile of ring surface mass density specified above. The local AMF is then converted into ring viscosity using Eq. (6).

Since we do not know the actual confining torque from Janus at the A ring edge, we consider a range of torques for the 7:6 ILR with the minimum value close to zero (very small phase lag), up to the maximum value, $T_R = 3.4 \times 10^{12}$ Nm. Figure 1 shows the resulting upper and lower limits on the AMF and the inferred viscosity as a function of radius across the A ring, As noted above, the small phase lag observed in the 7-lobed pattern indicates that the actual numbers are probably closer to the lower limit.

As might be expected, this plot shows a gradual outward decrease in the AMF (Fig. 1a) through the ring as we encounter successive resonances with the external satellites. The notable drops in flux at ~125,000 km, 130,500 km, 134,500 km, and 136,700 km are due to the Janus 4:3, 5:4, 6:5 and 7:6 ILRs, respectively, and the drop at ~132,300 km is due to the Mimas 5:3 ILR. While these resonances play a major role in removing the viscous flux of the A ring, resonances from the other satellites cumulatively have almost the same contribution, especially in the region outside the Encke gap where hundreds of first and second order resonances of Epimetheus, Prometheus, Pandora and Atlas accumulate. Finally, the viscous flux drops low enough for the Janus 7:6 ILR torque to remove the remaining angular momentum and prevent the ring's outer edge from spreading. In fact, our calculations indicate that the overlapping resonance torque from Atlas would be of the same order as that from the Janus 7:6 ILR if the phase lag of the 7-lobed pattern Δ~0.9°. Since the phase lag is indeed believed to be small, both torques may be involved in removing the remaining AMF at the ring edge.

Independent estimates of the ring's viscous AMF at the radii of the Encke and Keeler gaps are provided by the shepherding torques exerted by Pan and Daphnis, respectively, as indicated by the filled circles in Fig. 1a (with 3-sigma error bars on the satellite's masses taken from Weiss et al. 2009). It can be seen that the torque from Daphnis is quite consistent with our calculated curves. However, the shepherding torque from Pan is larger than our local estimate of the AMF. This may be because either our adopted surface mass density of the A ring is underestimated in the trans-Encke region by an average factor of 1.6 – 2.2 (if the surface mass densities were higher, than the torques from other satellite resonances here would be strong enough to bring the curve of the AMF near the Encke gap up to a level comparable to the torque from Pan), and/or the mass of Pan (from Weiss et al. 2009) is overestimated by a factor of 1.3 – 1.5 (as the torque goes as the square of the mass). Overall, our torque calculations appear to be accurate within a factor of 1.6 – 2.2, and most importantly are consistent with the observed decrease in the ring's AMF between the Encke gap and the edge of the A ring.

Although its net contribution to the AMF in the rings is small, Pan has an interesting relationship with the A ring. Not only it does it act to transfer angular momentum from one edge of the Encke gap to the other (via the overlapping resonances torque), it also transfers angular momentum on a larger scale as well. In the inner regions of the A ring, Pan decreases the AMF of the ring through its many ILRs, while in the trans-Encke region Pan increases it again through its almost-equally-numerous OLRs. The cumulative ILR torques slightly exceed that of the OLRs, partly because of the higher surface density in the inner regions of the A ring. (The impact on Pan's orbit is discussed in Section 4). A similar phenomenon occurs with Daphnis and the Keeler gap, but its overall effect on the A ring is negligible because of the small size of the discrete-resonance Daphnis torques. a

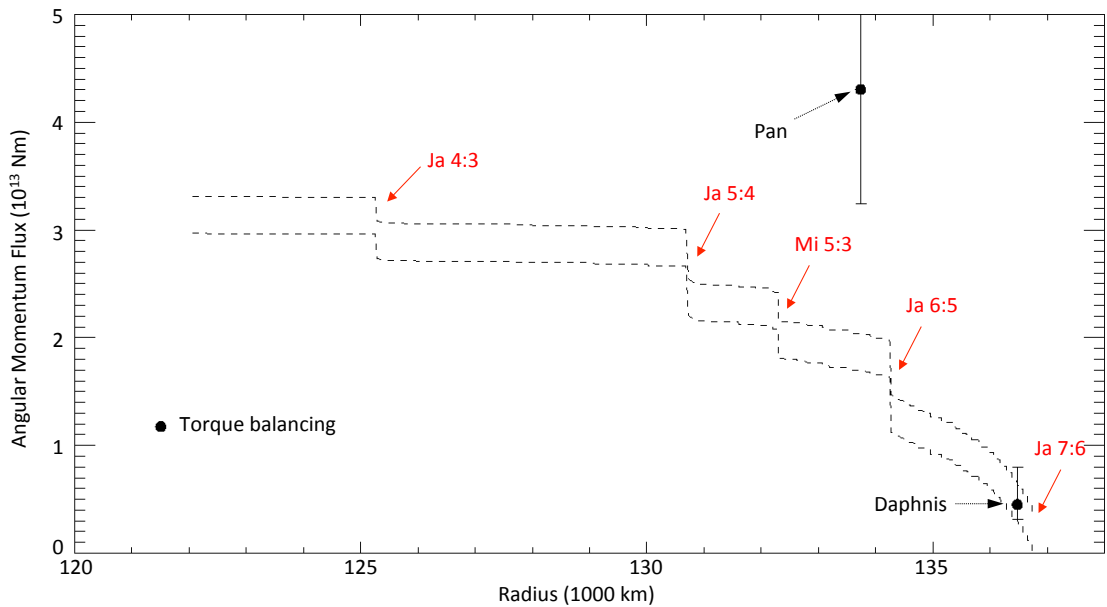

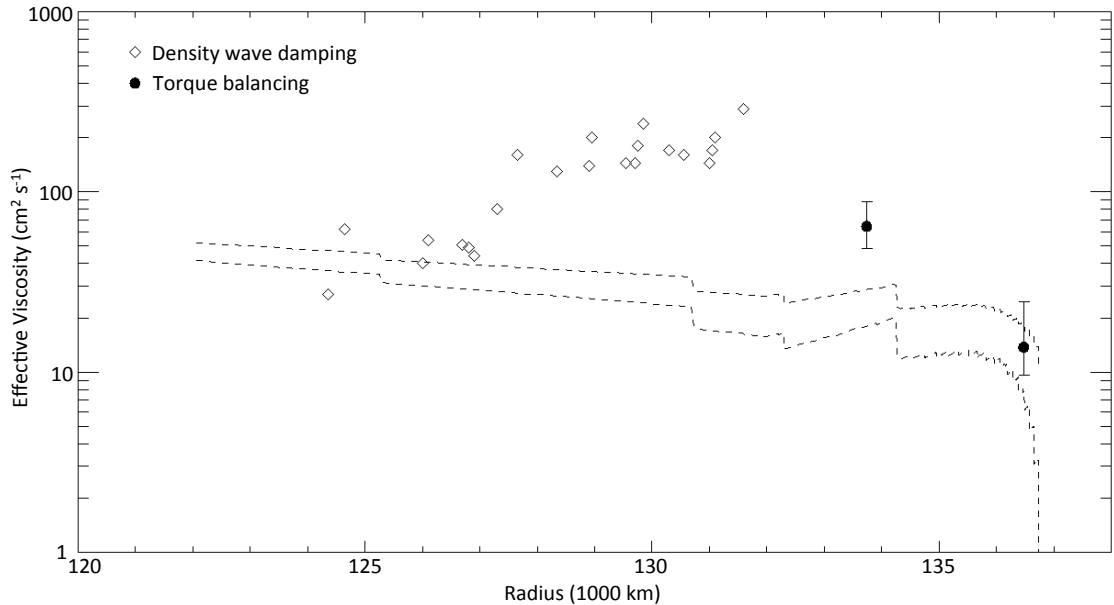

Figure 1. Upper and lower limit estimates (dashed lines) of (a) the angular momentum flux (AMF) and (b) the effective viscosity (on a log scale) in the A ring as a function of radius, taking into account all the first and second order Lindblad resonances from the satellites (Table 1) and the 5:3 bending wave from Mimas. The curves have upper and lower limits because the exact rate of angular momentum transfer from the Janus 7:6 ILR is uncertain (see text). Solid circles represent the AMF and the effective viscosity estimates derived independently from torque-balancing arguments at the Encke and Keeler gaps. Diamonds represent the viscosity of the ring inferred from the damping of density waves (Tiscareno et al. 2007); though there is no reason to believe that the viscosity inferred in this way should be the same as the effective viscosity estimated in this work (see text). Only the strong ILRs are identified in part (a) but many smaller ILRs contribute to the decrease of the AMF of the ring, especially beyond 130,000 km radius.

Figure 1b shows the effective viscosity inferred from Eq. (6) as well as that estimated above from balancing the shepherding torques at the Encke and Keeler gaps. Also shown here are the effective viscosity estimates inferred from damping of weak density waves (Tiscareno et al. 2007). Similar to the flux profile, the effective viscosity decreases in several small steps from the inner edge of the A ring up to the Encke gap, and then drops rapidly due to the large number of resonances in the region beyond this gap. However, unlike the viscous flux profile, the effective viscosity *increases* in some regions with increasing radius; it also decreases gradually when there are no resonances perturbing the ring. In both instances, the reason is that we calculate the effective viscosity at a certain radius by dividing the AMF of the ring (at that same radius) by $3\pi\Sigma na^2$ (Eq. 6). Hence, when the surface mass density $\Sigma$ increases with radius (for 122,000 km < $a$ < 132,000 km) the effective viscosity decreases, while when $\Sigma$ decreases with radius (for $a$ > 132,000 km) the effective viscosity can increase, unless the satellite torques are strong enough to decrease it. As a result of this relation between the surface density of the ring and the viscous flux the effective viscosity (both upper and lower limits) in the A ring decreases from 40-53 cm$^2$ s$^{-1}$ at the inner edge of the A ring down to 12-25 cm$^2$ s$^{-1}$ at a radius of 132,000 km, and then varies within a range of 15-30 cm$^2$ s$^{-1}$ between 132,000 and 136,000 km, before rapidly decreasing to less than 11 cm$^2$ s$^{-1}$ at the outer edge of the A ring.

Again, the calculated effective viscosity at the Keeler gap is consistent with the estimate obtained in Section 3.1 from torque balancing, but not at the Encke gap, probably for the reasons discussed above. Nevertheless, the general decrease of the effective viscosity in the trans-Encke region is consistent with that estimated from torque balancing.

Our effective viscosity estimates are also consistent with the viscosity inferred from the damping of density waves (Tiscareno et al. 2007) in the inner part of the A ring. However, they diverge from the density-wave-derived values in the outer part of the A ring, an issue to which we will return to in Section 5.

**3.3 Resonant confinement of the B ring**

Since our estimates of the A ring's AMF and its viscosity seem to be roughly consistent with those obtained from balancing the shepherding torques, we have also attempted to apply the same method to the B ring[2]. Compared to the A ring, the number of resonances in the B ring is small; Table 1 shows that there are 27 first and second order ILRs, among which the Mimas 2:1 ILR (at the outer edge of the B ring, at a radius of 117,553 km) is by far the strongest. In addition to resonances of the satellites involved in the confinement of the A ring, the B ring contains an Enceladus 3:1 ILR (radius ~ 115,207 km). The B ring has also a Mimas 4:2 IVR located ~830 km interior to the 2:1 ILR, at a radius of 116,724 km.

Unlike the situation in the A ring, the density and bending waves driven by these resonances are not easy to see in either imaging or occultation data. This is due to the large amount of confusing, small-scale structure in the B ring, as well as to its typically very high optical depth. In Voyager data, only the Janus 2:1 density wave and the Mimas 4:2 bending wave were detected (Esposito et al. 1983, Lissauer 1985). Recently however, Hedman and Nicholson (2016) have detected density waves due to the Mimas 5:2, Pandora 3:2, Enceladus 3:1, and Janus 3:2 ILRs in stellar occultation profiles obtained by Cassini.

Starting from the Mimas 2:1 ILR and moving inwards, we estimate the cumulative AMF across the B ring by taking into account the satellite torques from the expected ILRs (Table 1) and the Mimas 4:2 IVR. Hedman and Nicholson (2016) found that the surface mass density is surprisingly constant across the different regions of the B ring, ranging between 50 g cm$^{-2}$ and 75 g cm$^{-2}$ except for a single value of 120-140 g cm$^{-2}$ in the region around 116,100 km. For simplicity, we use a constant value of $\Sigma = 75$ g cm$^{-2}$ to estimate the satellite torques and thus the AMF in the B ring (the variation of the surface mass density at the outer edge of the B ring is more complex; see Longaretti 2017b).

---

[2] We do not study the Cassini division because of the existence of numerous complex structures in this region that remain to this day largely unexplained. Cassini has not found satellites in any of its narrow gaps (Nicholson et al. 2017), and there are relatively few and weak external satellite resonances. The major contributors to this region's AMF are thus currently unknown.

Unlike the outer edge of the A ring, a phase lag of $\Delta = 2.9^{\circ}$ in longitude between the $m=2$ forced pattern and the longitude of Mimas has been reported by Spitale and Porco (2010) based on the analysis of a large set of Cassini images. Nicholson et al. (2014), however, did not detect any significant lag in their analysis of radio and stellar occultation data, setting an upper limit of $1.8^{\circ}$. We thus prefer to treat the reported phase lag as an upper limit rather than as a strong constraint. Longaretti (2017b) shows that the torque expression at the sharp edge of the B ring from Borderies et al. (1982) is underestimated by up to a factor of ~6. We take this correction into account in calculating the resonant torque at the edge of the B ring and set this value (as well as the upper limit on the phase lag) as the starting condition for our calculation in Fig. 2a.

The variation of the viscous AMF across the B ring (Fig. 2a) is very different from that of the A ring. It remains relatively constant, decreasing by ~10% before dropping abruptly to zero at the Mimas 2:1 ILR. Apart from the latter, the largest torques are again associated with the two first-order ILRs of Janus, as indicated in the figure. Also unlike the situation in the A ring, where the Janus 7:6 ILR is only able to confine the outer edge of the ring with the help of many other resonances, application of Eq. (4) to Mimas' 2:1 ILR shows that its maximum torque could confine a ring (without the assistance of other satellite resonances) with a viscous AMF more than ten times larger than our best estimate for the B ring.

The relatively flat profile of the AMF also applies to the effective viscosity estimates (Fig. 2b); $\nu$ decreases from ~30 cm$^2$ s$^{-1}$ at the inner edge of the B ring to ~8 cm$^2$ s$^{-1}$ at the outer edge. Note the gradual outward decrease in the ring effective viscosity, even though we are assuming a constant value of surface density across the B ring. Here the denominator in the effective viscosity calculation is proportional to $a^{1/2}$ (see the discussion above) and therefore, for a constant viscous flux, the effective viscosity must decrease with increasing radius. Note that Fig. 2b shows the effective viscosity profile based on the measured phase lag in the $m = 2$ pattern at the outer edge of the B ring. But if this phase lag were much larger, then the torque from the Mimas 2:1 ILR alone would be capable of maintaining the edge of the B ring for an effective ring viscosity up to ~330 cm$^2$ s$^{-1}$.

Unfortunately and unlike the situation in the A ring, there are no satellite-controlled gaps within the B ring, which prevents us from making an independent check of our estimates of the AMF or the effective viscosity across the B ring.

Two conclusions can be drawn for the B ring: (1) unlike the Janus 7:6 ILR at the outer edge of the A ring, which depends on the help of many other resonances to confine this ring, the Mimas 2:1 ILR alone can easily prevent the outer edge of the B ring from spreading outwards, and (2) the effective viscosity is expected to be relatively constant across the ring, perhaps ranging between 24 cm$^2$ s$^{-1}$ and 30 cm$^2$ s$^{-1}$.

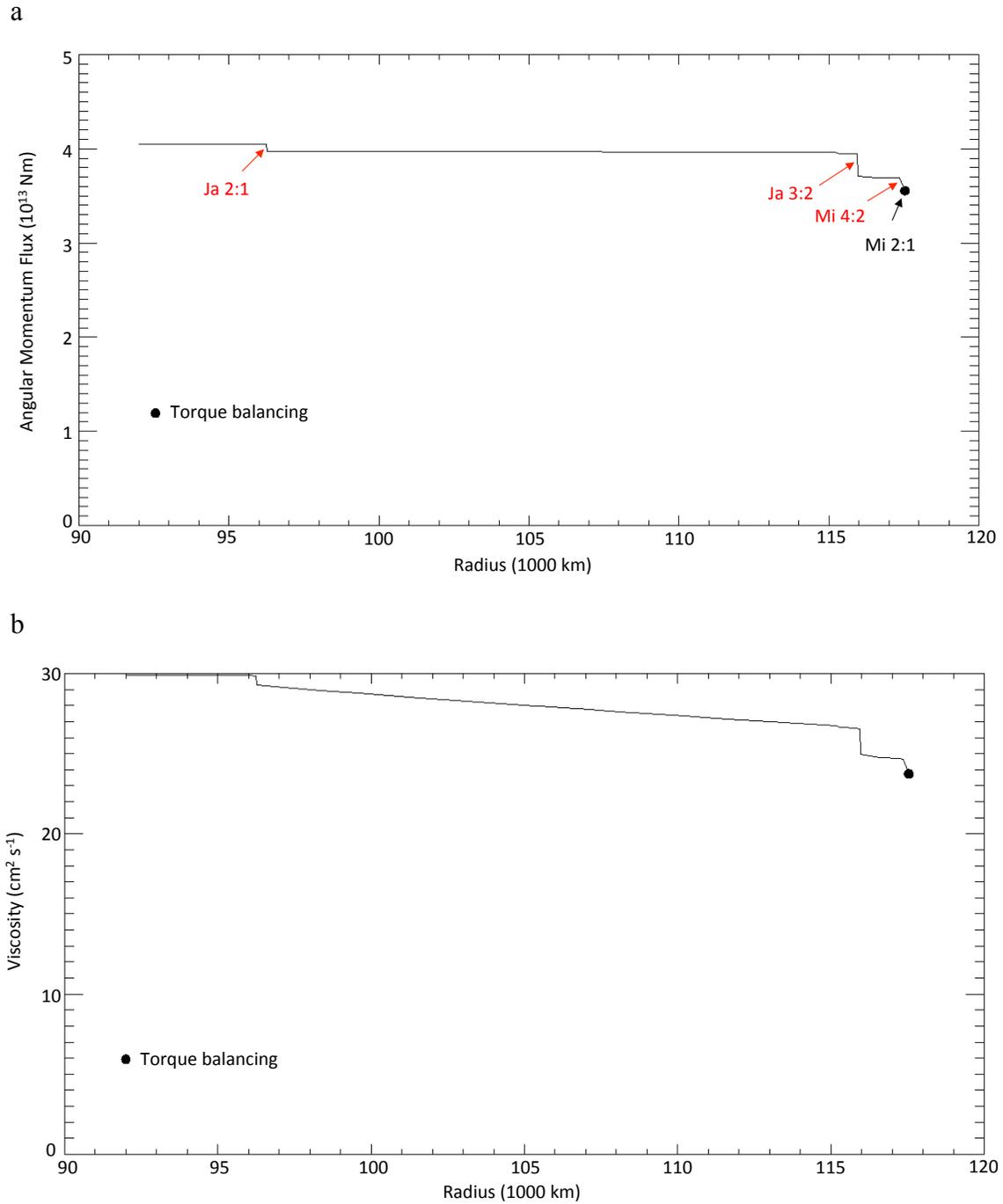

Figure 2. Upper limit estimates of the angular momentum flux (AMF) (a) and the effective viscosity (b) in the B ring as a function of the radius, taking into account all the first and second order Lindblad resonances from the satellites (Table 2) and the 4:2 bending wave from Mimas. The solid circle represents the AMF, and the resulting effective viscosity, estimated from torque balancing at the outer B ring edge with the reported phase lag relative to Mimas of $\Delta = 2.9^\circ$.

## 4. Evolution of the satellites of Saturn

So far we have studied the role that the satellites play in the modification of the AMF and the viscosity of the rings through various Lindblad and vertical resonances. The rings, on the other hand, respond to these torques by exerting an equal and opposite reaction torque on the satellite at each resonant location. Such a response has consequences for the evolution of the satellites' semi-major axes.

We sum the torques exerted by the A and B rings on each of the satellites of Saturn, as calculated above, and compare them to the torques exerted by Saturn's tidal bulge using the standard expression (Kaula 1964),

$$T_{Tides} = \frac{3}{2}\frac{k_2}{Q}GM_s^2\frac{R^5}{a_s^6}. \qquad (10)$$

Here $k_2$ is Saturn's tidal Love number that represents the planet's reaction to the tidal potential due to a satellite and $Q$ is the tidal quality factor that models the phase lag between the planet's tidal bulge and the direction of the satellite. To calculate the planet's tidal torque on the satellites we use the value of $k_2/Q = 1.59 \times 10^{-4}$ that was empirically determined by Lainey et al. (2012, 2017) using Earth-based and thousands of Cassini-based astrometric positions of the satellites of Saturn (Tajeddine et al. 2013, 2015; Cooper et al. 2014). In addition, we calculate the rate of change of the satellites' semi-major axes due to the torques from the planet and the rings (Kaula 1964),

$$\left.\frac{da}{dt}\right|_{Tides} = 3\frac{k_2}{Q}\frac{M_s}{M_p}\frac{R^5}{a_s^4}n_s \qquad (11)$$

and

$$\left.\frac{da}{dt}\right|_{Rings} = \frac{2a_s^{1/2}}{M_s(GM_p)^{1/2}}T_{Rings} \qquad (12)$$

where $T_{Rings}$ is the total torque from the rings on the satellite. Table 2 shows the torques and the resulting rates of change of the semi-major axis for different satellites. Interestingly, the ring torque dominates the orbital evolution of the semi-major axis for all the small, inner satellites of Saturn. On the other hand the ring torque on Mimas is ~10% of that due to Saturn's tides.

The numbers from Table 2 show that the torque from the rings should be taken into account when modeling the orbital evolution of all the "ring moons" of Saturn as well as Mimas. However, the ring torque becomes negligible in comparison to the tidal torque for satellites beyond Mimas because they have no first order ILRs (and some of them have only a very small number of second order ILRs) in the rings. Of course our calculations do not take into account other effects that can play an important role in the satellites' orbital evolution, in particular the gravitational interactions between satellites. For example, Janus and Epimetheus are locked together in mutual horseshoe orbits, while Prometheus and Pandora are involved in overlapping mean motion resonances that put them on chaotic orbits within the resonant region (Goldreich & Rappaport 2003). Atlas is currently locked in a 54:53 mean motion resonance with Prometheus (Cooper et al. 2015).

Subject to the above caveats, we can estimate the outward migration time scale for the satellites, defined as $\Delta t = \Delta a / \dot{a}$, where $\Delta a$ is the distance between the satellite and the edge of the A ring and $\dot{a}$ is the average rate of change of the semi-major axis of the satellite due to torques from the rings and Saturn's tides from Table 2. Of course this is only a rough calculation, inasmuch as we are assuming a constant value of $\dot{a}$, which will in reality be higher when the satellite is closer to the rings and thus has more ILRs within the A ring. Our estimates therefore represent *upper limits* on the ages of the satellites.

We find maximum ages on the order of $\Delta t_{Mimas} \sim 10^9$ years, $\Delta t_{Epimetheus} \sim 10^9$ years, $\Delta t_{Janus} \sim 3 \cdot 10^8$ years, $\Delta t_{Pandora} \sim 10^8$ years, $\Delta t_{Prometheus} \sim 10^7$ years, and $\Delta t_{Atlas} \sim 4 \cdot 10^6$ years (we assume that Atlas was trapped in resonance with Prometheus only recently). All these satellites thus appear to be younger than 1 Gyr. These calculations do not take into account the inter-satellite interactions: Mimas and Tethys are in a 4:2 mean motion resonance. Eqns. (10) and (11) give for Tethys $T_{Tides} = 7.41 \cdot 10^{15}$ Nm and $\dot{a}_T = 66.8$ km/Myr, and since they are locked in a resonance, the actual migration rate of Mimas due to tides would be $\dot{a}_M = (n_1/n_2)^{-2/3} \dot{a}_T = 42.1$ km/Myr; thus, Tethys may be decelerating Mimas' orbital migration, and if the latter has been caught in the resonance since its formation, the migration time-scale would then be slightly larger. However, recent studies of Mimas suggest that it may possess a global subsurface ocean (Tajeddine et al. 2014, Noyelles et al. 2016, Noyelles 2017, Caudal 2017), although such a claim is difficult to reconcile with the absence of surface fracturing (Rhoden et al. 2017). If this is true, orbital migration could have been delayed recently by tidal dissipation in the satellite, thus increasing the estimated age.

The Mimas-Tethys reasoning may also apply to the Janus/Epimetheus pair, although these two satellites may have entered into their mutual horseshoe libration quite recently, since the lifetime of such a configuration is estimated to be only $\sim 2 \cdot 10^7$ years (Lissauer et al. 1985).

The upper limits on the ages of Prometheus and Pandora that we obtain are larger than those of Borderies et al. (1984), who estimated $\sim 10^6$ and $10^7$ years, respectively. This is probably due to differences in the input parameters; for instance, they used a ring surface mass density of 50 g/cm$^2$ (a factor of $\sim 3$ larger than what is used here). It is also unclear what satellite masses they used, but if they assumed a satellite density of 0.92 g/cm$^3$ for ice (which is twice what Cassini measured; Thomas 2010), this could account for more of the difference. It is also unclear whether their age estimate was based on a numerical integration over time, or a simple instantaneous estimate like that in Table 2.

Borderies et al. (1984) also raised the question of whether Pandora will eventually be captured into the 3:2 ILR with Mimas (radius $\sim 141,892$ km), which is currently just beyond Pandora's orbit (mean radius $\sim 141,710$ km). Furthermore, Pandora is currently located between inclination type (radius $\sim 141,532$ km) and eccentricity type (radius $\sim 141,827$ km, $\sim 68$ km inside the 3:2 resonance) 6:4 Mimas resonances and, surprisingly, was not caught in either of them in the past (Murray & Dermott, 1999, see their Fig. 10.28). The rate of orbital migration of Mimas is $\sim 42.1$ km/Myr (taking into account the resonance with Tethys), but the Mimas 3:2 ILR moves more slowly, at a rate $(3/2)^{-2/3} \dot{a}_M$, or $0.76 \dot{a}_M \approx 32$ km/Myr. Close to that of Pandora ($\sim 45$ km/Myr taking into account the uncertainties in the torques exerted by the A and B rings, and Saturn's $k_2/Q$), meaning that the latter may have managed to avoid getting caught in the Mimas resonances by migrating at the same rate as the resonances surrounding it. On the other hand, if we assume that the values in Table 2 are accurate, then the Mimas 6:4 inclination type resonance would catch up with Pandora in another ~15 Myr, of the same order as the age of Pandora. This number may, however, vary depending on the Saturn's $k_2/Q$ at Mimas' tidal frequency.

The migration times of the inner moons and Mimas confirms that their ages are less than 1 Gyr, more consistent with a model of satellite formation from the rings (Charnoz et al. 2010, 2011), than with a primordial origin. We have, however, not attempted to construct a realistic history of the satellites orbits, because this would involve a self-consistent reconstruction of which resonances fell within the rings at any given time, as well as assumptions as to which resonance(s) may have terminated the A ring at various points during this evolution. A further complication is the possibility that the present ring moons may represent merger products from an initially much larger set of smaller objects that accreted at the edge of the A ring (Charnoz et al. 2010).

Furthermore, note that we used here a constant value of the effective $k_2/Q = 1.59 \times 10^{-4}$ (Lainey et al. 2012, 2017), hence an effective quality factor of $Q \sim 1600$, ten times smaller than the commonly accepted value of ~16,000 (Dermott et al. 1988) based on analysis of past resonances. Lainey et al. also found that the quality factor is even smaller at the tidal frequency of Rhea with $Q\sim300$. Internal oscillation modes inside Saturn raised by the satellites (Ogilvie & Lin 2004) are believed to be the cause of such small values of the effective $Q$ (Fuller et al. 2016). Therefore, it is unknown when Mimas got locked in those resonances and started migrating rapidly. If this happened soon after its formation, then our age estimates are correct; but if it happened only recently then Mimas may be as old as the solar system. On the other hand, the differences in the tidal quality factor do not affect the age estimates of the inner satellites since their migration rates are dominated by ring torques (Table 2).

Pan constitutes a particularly interesting case; it has ILRs in the inner part of the ring but because it is within the A ring, the reaction torques from these are largely countered by those from the OLRs in the trans-Encke region. However, Pan has more ILRs than OLRs in the rings (see Table 1), and the surface mass density $\Sigma$ is larger interior to the Encke gap. Therefore, (as Table 2 shows) the net response from the rings is a positive total torque on Pan, making it want to drift slowly outward. However, the situation is more complex than this. If we calculate the ratio between the angular momentum of the ring beyond the Encke gap and that of Pan, we find that $|L_{Ring}/L_{Pan}| = 4.7\times10^3$. This means that although the discrete resonant torques on Pan tend to push it outward, Pan cannot easily drag the nearby ring material with it. Instead the much larger shepherding torques at the gap edges push Pan back to the middle of the gap, or perhaps a little outside this at the point of overall torque balance.

We find a similar situation for Daphnis and the Keeler gap, but further complicated by the presence of a first-order Prometheus ILR at the gap's inner edge (Tajeddine et al 2017). It thus seems likely that these two satellites will only exit the rings when the satellites holding back the A ring have migrated far enough to allow the ring to spread beyond the Roche limit, dropping the ring's surface mass density, and eventually converting some of the outermost ring material into new satellites. But these new satellites could confine the ring again, further delaying the exit of Daphnis and Pan from their prisons.

| Satellite | Planet's Tidal Torque (Nm) | Ring Torque (Nm) | $\dot{a}|_{Tidal}$ (km/Myr) | $\dot{a}|_{Rings}$ (km/Myr) |
|---|---|---|---|---|
| Mimas | $3.90 \times 10^{14}$ | $3.97 \times 10^{13}$ | 46.4 | 4.72 |
| Epimetheus | $2.92 \times 10^{11}$ | $1.43 \times 10^{12}$ | 2.21 | 10.8 |
| Janus | $3.80 \times 10^{12}$ | $1.89 \times 10^{13}$ | 7.99 | 39.5 |
| Pandora | $2.95 \times 10^{10}$ | $1.57 \times 10^{12}$ | 0.83 | 44.2 |
| Prometheus | $4.41 \times 10^{10}$ | $1.14 \times 10^{13}$ | 1.06 | 274 |
| Atlas | $8.14 \times 10^{7}$ | $4.05 \times 10^{11}$ | 0.047 | 233 |
| Pan | $4.09 \times 10^{7}$ | $5.12 \times 10^{11}$ | 0.036 | 90.1 |

Table 2. Comparisons between the torques exerted by the rings with those by Saturn's tidal bulges on the inner satellites and Mimas, with the resulting migration rate of their semi-major axes due to each effect. Note that the latter ignore any resonant interactions between the satellites (see text).

## 5. Discussion

Our calculations in Section 3, illustrated in the upper panels of Figs. 1 and 2 strongly suggest that the AMF transported outwards though the rings is almost constant across the B ring, but decreases significantly across the A ring, especially in the trans-Encke region. This is a direct consequence of the high concentration of first-order satellite resonances in this region, and their relative paucity in the B ring, and seems to be a robust result. A corollary to this result is that the confinement of the A ring is best thought of as distributed across many satellite resonances of comparable strength, rather than being due to a single ILR at the ring's outer edge.

But simply concluding that the AMF must decrease across the A ring does not explain *why* this happens; nor does it explain why the viscosity estimates for the A ring differ so much between this work and those inferred from the damping of density waves (Tiscareno et al. 2007). These points are addressed in this section, starting with the issue of viscosity.

### 5.1 Process dependent viscosity

The viscosities in the A ring obtained from Eq. (6) and those inferred from density wave damping (Fig. 1b) are different because they are inferred from different physical processes and contexts. Several points must be kept in mind when trying to relate different measurements of viscosity: first, different levels of nonlinearity will lead to different effective viscosities; second, in the nonlinear regime, no single effective viscosity can be defined even for a given process (e.g., collisional stresses); and last, different processes depend on different effective stress tensors.

*5.1.1 Collisional stress*

Collisions are collectively characterized by the collisional pressure tensor $P_{ij}$ ($i,j = x,y,z$ or $r,\theta,z$). The AMF $\propto a_{r\theta}$, the energy dissipated $\propto a_{r\theta} - 2qt_1/3$ and the damping of eccentricities (e.g. in density waves) $\propto t_1$, where $a_{r\theta}$ is the azimuthal average of $P_{r\theta}$ while $t_1$ is an average over a combination of $P_{rr}$ and $P_{r\theta}$ (Borderies et al. 1985, Shu et al 1985). The quantity $q$ characterizes the compression of the flow, and has two contributions, one from the gradient of eccentricity and the other from the gradient of apsidal shift of the ring streamlines; more formally, using the notation of Eq. (A1), $q = [(ade/da)^2 + (maed\Delta/da)^2]^{1/2}$ (see, e.g., Longaretti 2017a and references therein for more details). One may always write[3] $a_{r\theta} = S_0 \sigma_0 \nu_0 G(q)$ and $t_1 = -\Omega \sigma_0 \nu'_0 F(q)$ ($S_0 = 3\Omega/2$ is the unperturbed velocity shear) but as $F$ and $G$ are different functions of $q$, the effective 'transport' and 'dissipation' viscosities ($\nu_t = \nu_0 G(q)$ and $\nu_d = \nu'_0 F(q)$ respectively) will differ, possibly in a significant way, especially in the AMF reversal regime. This is true even though density wave damping is modeled in the linear limit, as the measured viscosity will then incorporate nonlinear effects. (Note that in a circular flow, $q = 0$, the damping term $t_1$ cancels, and the dissipation ($a_{r\theta} - 2qt_1/3$) and transport ($a_{r\theta}$) terms become identical so only one effective viscosity is relevant in this limit.) Note also that the point made above about effective transport and dissipation viscosities in the nonlinear regime is also valid for a Newtonian fluid, although the microscopic kinematic viscosity $\nu_0$ is unique and unambiguous in this case.

This short discussion justifies the first two statements above for collisional stresses. Let us now turn to other types of stresses.

*5.1.2 Self-gravity wakes stress*

It is well-known that such wakes produce a self-gravitational stress tensor, whose contribution to the vertically averaged fluid momentum equation is of the form $(1/\sigma)(\partial T_{ij}/\partial x_i)$ with $T_{ij} = \int dz\, (g_i g_j - g_i^2 \delta_{ij})/4\pi G$ (in cgs units) and where $g_i$ is the self-gravitational acceleration in the $i$ direction (Lynden-Bell and Kalnajs 1972). This is formally equivalent to the pressure tensor term $(1/\sigma)(\partial P_{ij}/\partial x_i)$. In practice, self-gravity wakes fluctuate on short length and time scales (comparable to the critical wavelength of the instability and to the associated growth time scale). It is common in fluid dynamics to average such fluctuations over such scales in order to obtain an effective laminar flow equation, with an extra averaged self-gravitational stress tensor term $<T_{ij}> = \int dz <g_i g_j - g_i^2 \delta_{ij}>$. Although $<g_i> = 0$, this averaged self-gravitational tensor does not vanish because fluctuations correlate across different components. This averaged stess tensor has the same physical dimensions as the pressure tensor so that one may always define as before effective 'transport' and 'dissipation' viscosities associated with self-gravity wakes, by substituting the averaged self-gravity tensor for the pressure tensor into the corresponding quantities $a_{r\theta}$ and $t_1$, which will then possess a self-gravitational component on top of the collisional one.

---

[3] Two different kinematic viscosities $\nu_0$ and $\nu'_0$ have been introduced to stress the fact that in a non-Newtonian fluid, there is no reason that a single such quantity may be defined, although the difference could be absorbed in the definition of $F$ and $G$.

The behavior of this self-gravity tensor is not known except for the angular momentum transport contribution $<T_{r\vartheta}>$ in a circular Keplerian background flow (Daisaka et al., 2001); in particular, it's value is not known in perturbed flows. However, one may expect the remarks made for the collisional stress tensor to apply here as well: the self-gravitational stress will depend on the nonlinearity parameter $q$, in an unknown way, but there is no reason to expect its effective transport and dissipation components to be of the same magnitude.

*5.1.3 "Reynolds" stress*

This discussion would not be complete without mentioning the third tensor of interest, $R_{ij} = <\delta(\sigma v_i v_j)>$ whose contribution to the effective laminar flow $(1/\sigma)(\partial R_{ij}/\partial x_i)$ is formally identical to that of the collisional and self-gravity stresses ($v_i, v_j$ stand for the velocity and $\sigma v_i v_j$ is the deviation of the tensor from its laminar large scale form). This tensor is a generalized form of the usual Reynolds tensor which takes into account surface density fluctuations. Self-gravity wakes do drive fluctuations in density and velocity along with fluctuations in self-gravity, and Daisaka & Ida (1999) have shown that the contribution to this tensor from self-gravity wakes is comparable to the self-gravitational tensor. All the remarks made above concerning the self-gravitational stress apply to this generalized Reynolds stress as well.

*5.1.4 General stress tensor and its implications*

One can therefore define a generalized stress tensor $W_{ij} = P_{ij} + T_{ij} + R_{ij}$. Correlatively the form of $a_{r\theta}$ and $t_1$ associated with $W$ contains contributions from all three tensors discussed above, which contribute to angular momentum transport for the first and density wave damping for the second, with three different dissipation and transport effective viscosities[4] and related $F$ and $G$ dimensionless functions[5] of $q$.

With these remarks in mind, let us return to the problem at hand. In the A and B rings the known processes that control the effective transport and dissipation viscosities just defined are either collisions, self-gravity wakes, or small-scale viscous overstabilities. In unperturbed regions of the A ring, it is nearly certain that angular momentum transport is dominated by self-gravity wakes[6]. The situation is less clear in perturbed regions, but one may question the existence of small scale self-gravity wakes in density waves on two fronts: first, the typical growth time-scale of wakes is comparable to the orbital period, i.e., to the peak-trough travel time of the flow; second, the ring density in the peaks may be sufficient to quench the wake instability. Therefore it is not clear that the process dominating density wave damping is the same as the transport process in unperturbed regions in the A ring. Furthermore, as the preceding discussion has shown, there is also no reason, for a given process, to expect that the effective damping viscosity of density waves has any simple relation to the effective transport viscosity in unperturbed regions of the A ring. Consequently, there are no grounds to expect that the damping viscosity measured in density waves should bear any simple relation to the transport viscosity derived in this work.

---

[4] In principle, other processes can contribute to the generalized Reynolds stress tensor, e.g., small-scale viscous overstabilities. This process has only been studied for axisymmetric modes, but nothing prevents in principle the existence of non-axisymmetric small-scale disturbances. Although such terms are known to be stable in the shearing sheet limit (Schmidt et al 2001), they may be transiently excited and contribute to the problem. In the absence of other bulk forces, no other relevant stress tensor may be defined.

[5] From a conceptual point of view, such a dimensionless function must depend on all the dimensionless numbers characterizing the flow, in order for the associated viscosities $\nu_0$ to be a constant. For example, it is well known that the collisional stress tensor also depends on the effective optical depth.

**5.2 Density wave overlap and angular momentum transport reversal.**

Let us now turn to the question of why the AMF decreases so dramatically across the A ring, especially in the trans-Encke region. We propose that this is due to the phenomenon of angular momentum flux reversal (see below) on a large scale, associated with the large number of density waves in the outer A ring.

Eq. (6) is commonly used under the assumption that the radial gradient in angular velocity across the rings is described by simple Keplerian shear between nearby circular orbits: $dn/da = -(3/2)\, n/a$; i.e., $G(q)=1$ is assumed. This is violated in perturbed regions for the collisional stress tensor. In fact, if $q$ reaches a critical value of $q_a \sim 0.7 - 0.8$ then both the radial shear and the AMF can go to zero. This phenomenon --- sometimes referred to as "flux reversal" as the AMF can actually change sign if $q > q_a$ --- is thought to play a key role in the shepherding of narrow ringlets and in the maintenance of extremely sharp edges controlled by external resonances.

Indeed, angular momentum transfer to external satellites at a sharp ring edge requires either AMF reversal or a significant decrease of the surface density, when moving towards the edge. In proto-planetary disks, a decrease of the surface density over the whole perturbed region is the normal process of angular momentum transfer from the disk to an embedded planet (see, e.g., Crida et al, 2006). The fact that edges are sharp in planetary rings is quite unexpected in this respect, and requires the effective viscosity to go to zero if the surface density remains more or less constant (remember that the only factors in the AMF are the surface density and the effective viscosity, whatever the dominant transfer mechanism). This question was resolved by Borderies et al. (1982) for isolated resonances and Borderies et al. (1989) for overlapping ones; the role of AMF reversal was explicitly shown for collisional stresses in these two analyses.

---

[6] The length scale of small-scale overstabilities is comparable to that of self-gravity wakes, but their growth time is much slower so that the associated effective viscosity contribution to the Reynolds stress is expected to be much lower than for the wakes. Also the direct contribution of viscous overstabilities to the stress tensor terms is small. Consequently, viscous overstabilities contribute little to the various tensors previously introduced and are ignored for the rest of this discussion. Note that self-gravity wakes and viscous overstabilities do not seem to cohabit easily in numerical simulations (Salo et al, 2001).

In the present context, we suggest that the cumulative effect of the many density waves in the outer A ring is to produce an effective value of *q* which is sufficient to significantly reduce the AMF below its unperturbed value without a significant drop of the surface density over the same radial domain. To illustrate our argument, we plot in Fig. 3 the number of overlapping density waves as a function of radius in the A and B rings, averaged with a sliding mask of 100 km in width. For the purpose of this calculation, we consider a density wave to extend for a radial distance of six times its first wavelength. As this figure shows, the large number of density-wave-overlaps in the outer A ring indicates a highly perturbed region, with at least two waves co-existing at almost all radii outside 130,000 km.

This view does, however, raise two issues that we can only speculate upon considering the lack of relevant theoretical analyses in the literature:

1. The correct expressions for AMF are not certain for other small-scale transport processes besides collisional stresses.
2. Angular momentum is transported both at small scales (sub-laminar stresses) and large scales (density waves). A natural question here is the relative magnitude of these two contributions.

The second point does not really raise a problem of principle. Small and large-scale transports do actually occur at the same time. We have shown that each density wave removes only small fraction of the AMF in the A ring while collectively, they remove most of the AMF. The presence of density waves also serves as a background to produce some level of nonlinearity in the AMF process.

The first point calls for some comments. Note first that in practice, only self-gravity wakes need to be considered here (see the last footnote). Now, the turbulent viscosity prescription (which in effect is what is used here, although we deal with weak 2D instead of strong 3D turbulence) states that $< T_{r\vartheta} > \propto \sigma_0 \nu_0 S$ where *S* is the actual velocity shear (Prandtl 1925). The dependence on the shear (instead of, e.g., the rotational velocity) is in fact inevitable from a thermodynamic perspective (Longaretti, 2002; Lesur & Longaretti, 2005). A sheared medium is out of global thermodynamic equilibrium: the equilibrium distribution function of a gas whose entropy is maximized under the requirement of conserved angular momentum and energy exhibits uniform fluid rotation. Consequently, such a system will try to restore thermodynamic equilibrium by removing the shear, i.e., transporting angular momentum across the shear so that $< T_{r\vartheta} > \propto S$ seems necessary[7].

The point here is that wake transport *may* be proportional to the actual perturbed shear, a feature that can only be investigated through numerical simulations. On the other hand, the steady decrease of the AMF in the outer A ring region disclosed in this work requires either a slow decrease of the outer A ring surface density in the density wave overlap region, or a slow decrease of the angular momentum flux through flux reversal in the same region. As the first way is excluded from the known constraints on the surface density profile (the A ring edge *is* sharp), we are left with only two options in order for flux reversal to actually take place: either wakes are quenched or do not dominate the generalized stress tensor $W$ in density waves, or the wake self-gravitational stress does also exhibit AMF reversal. These two options are not mutually exclusive.

---

[7] In fact, this argument only proves that $< T_{r\vartheta} > \propto S^p$ where *p* > 0, but, on the one hand, experience in various contexts indicates that *p*=1 is the most common situation, and, on the other hand, a different power dependence does not change the major conclusion.

### 5.3 Implications and miscellaneous points

Let us now come back to the main thread of this discussion. According to this interpretation of our results, the AMF is indeed reduced in the trans-Encke region. This explanation has the benefit of causally connecting the decreasing AMF in the outer A ring with the concentration of resonances in this region. The high level of perturbation in the outer A ring (cf. Fig. 3a) is consistent with the steep decline of the ring's AMF in that region. However, in light of the preceding discussion, making this explanation quantitative is clearly beyond the scope of the present paper, as it involves sophisticated dynamical developments.

On the other hand, the number of resonance overlaps in the B ring is relatively low (see Fig. 3b) and, as suggested above, the outer edge is confined through flux reversal at a single resonance rather than via flux reversal due to multiple resonances within the ring.

Although our expressions for the satellite resonance torques, both for Lindblad and vertical resonances, are believed to be secure when the resonances drive density or bending waves in the rings, the application of these same expressions is less certain when the resonance falls at a sharp edge in the rings. The wave-derived torques provide an estimate of the *maximum* torque at a sharp edge (see Longaretti 2017b for detailed discussion of this point). This is a source of potential error in the "boundary conditions" we apply at the outer edges of the A and B rings, where we set the AMF equal to the torques associated with the Janus 7:6 and Mimas 2:1 ILRs, respectively.

It is also possible that there are additional sources or sinks of angular momentum within the rings besides satellite resonances. In particular, several density waves in Saturn's C ring have been identified as being driven by internal oscillations within the planet (Marley & Porco 1993, Hedman & Nicholson 2013, 2014), while other waves have been identified with tesseral resonances, i.e., permanent gravitational anomalies or slowly-propagating waves within Saturn (Hedman & Nicholson 2014, El Moutamid et al 2017). Many of these waves correspond to OLRs, at least in the C ring, and thus should lead to an outward *increase* in the AMF through the rings, unlike almost all satellite resonances. If similar waves exist in the A and B rings, something which has yet to be established, they might act to *reduce* the cumulative AMF somewhat below that shown in Figs. 1 and 2.

Ballistic transport constitutes another source of angular momentum exchange between rings and the outside world (Durisen et al. 1989, Estrada & Cuzzi 1996, Schmidt et al. 2009, Estrada et al. 2017). Ring particles in low optical depth regions of the ring acquire angular momentum from meteoritic bombardment, moving them further away from the planet. However, this process is less efficient in opaque rings. Hence, it is suspected to be the principle mechanism confining the inner edges of the A and B rings (by stripping material from the Cassini division and the C ring), and may even be the source of the AMF in the innermost regions of the A and B rings (cf. Figs. 1 & 2).

a

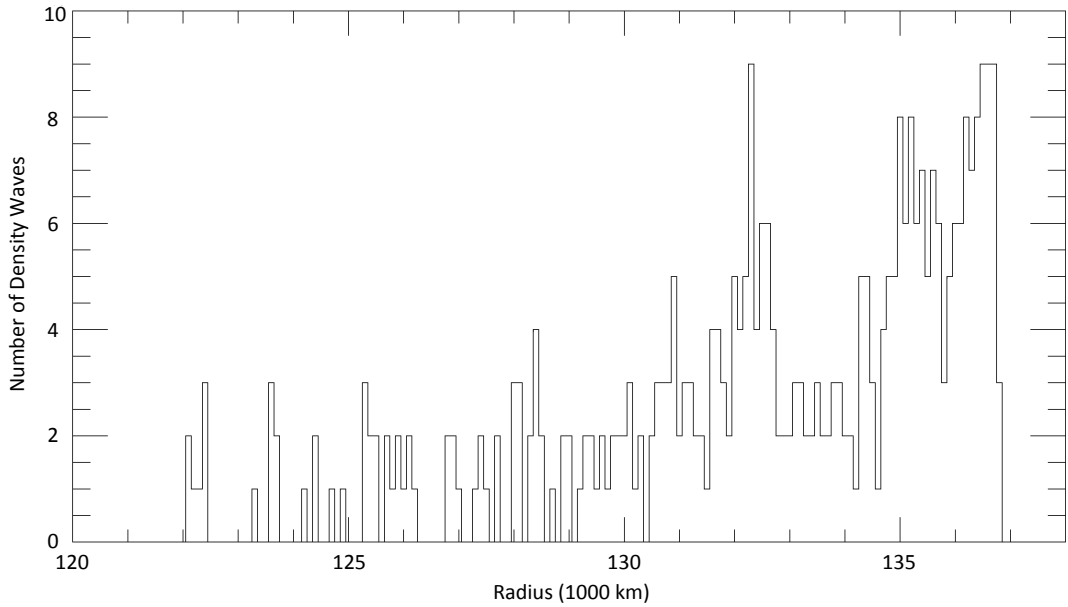

b

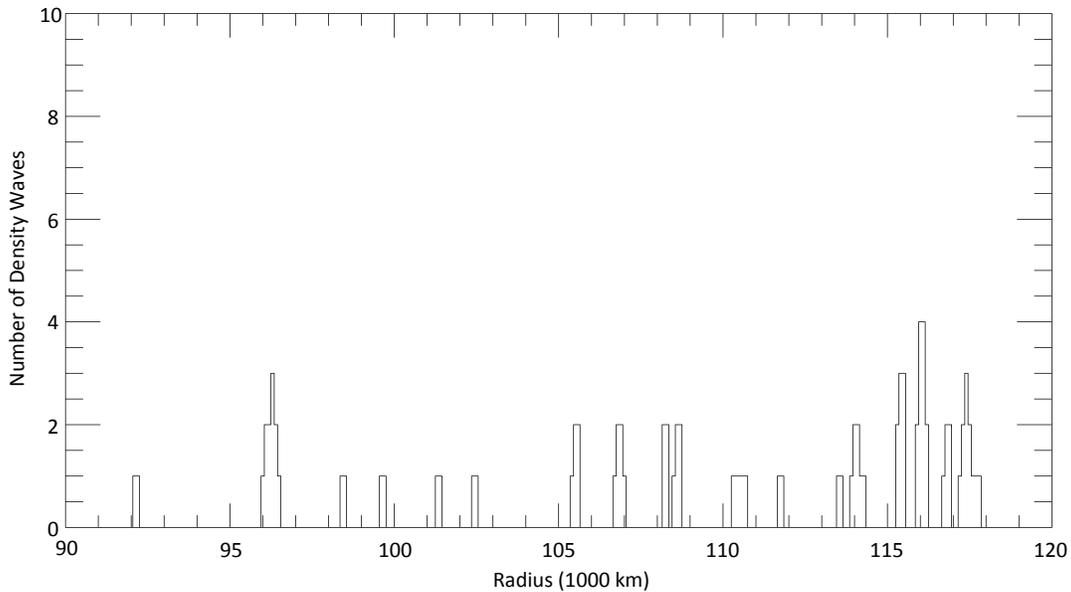

Figure 3. Number of density waves at any radial location in the A (a) and B (b) rings, averaged with a sliding mask of 100 km of radius. Here we limit the extension of a density wave to six times the first wavelength (calculated using Eq. 7.43 from Longaretti, 1992). While the B ring does not show much density wave overlaps, the A ring has a high number of overlaps near its outer edge, which supports the flux reversal mechanism to take effect.

## 6. Conclusion

A careful accounting of all known satellite resonances in the A and B rings reveals that, in the A ring, the radial confinement of the ring against viscous spreading is distributed over many resonances. As a result, the AMF is expected to decrease outwards across the ring. For the B ring, on the other hand, almost all of the work is done by the Mimas 2:1 ILR located at the outer edge. By equating the computed AMF with that expected to be transported by collisional interactions within the rings, we derive a radial profile of transport viscosity across both rings. In the A ring this is found to be roughly consistent with viscosities inferred from torque balancing at the Encke and Keeler gaps, but no such direct test is possible for the B ring. We speculate that the observed steep decrease in AMF in the trans-Encke region of the A ring may be due to "flux reversal" associated with the disturbed streamlines produced by the large concentration of density waves driven by satellite resonances in this region, which translates into a decrease in effective viscosity. Finally, we find that the reaction torques exerted by the rings on the small inner satellites, or "ringmoons", exceed the torques due to tides they raise in the planet, and lead to upper limits on the ages of the larger ring moons of 1 Gyr or less.

# Appendix A: Satellite resonant torques

Satellites can excite waves in planetary rings. The strongest such waves (density waves) arise from first order resonance motions excited by the equatorial, circular part of the satellite orbit. Radial (eccentric) oscillations of the satellites contribute higher order resonances and also produce density waves, but with weaker wave responses from the rings. Similarly, vertical oscillations induce vertical waves in the rings (bending waves). but only at second and higher order resonances.

The question of the angular momentum exchange (torques) between rings and satellites has long been scrutinized, starting with the original work by C.C. Lin, F. Shu, A. Toomre, P. Goldreich and S. Tremaine in the 1960s and 1970s. Such torques can in principle be computed in two different ways:

- From the angular momentum density carried by the wave at its group velocity (this results in a radial flux of angular momentum); or
- From a direct analysis of the torque itself (i.e., the coupling of the satellite forcing to the wave response).

The conservation of the so-called wave action ensures that, in the dissipationless limit, the overall flux carried by the wave (sometimes referred to as the angular momentum luminosity) is constant (independent of radius) and equal to the total satellite torque. This conservation of the angular momentum flux is also commonly used to compute the change in amplitude of the wave as it propagates.

Even though the torque is a second order quantity (with respect to the perturbing potential of the satellite, i.e., in the ratio of the satellite mass to the planet's), the direct method (the second way above) requires only a first order Lagrangian solution of the dynamics, though a second order solution is required if one uses Eulerian fluid dynamics. Conversely, the indirect method (the first way above) requires only first order Eulerian solutions. Eulerian fluid dynamics being much more widely used by theorists, most torque derivations rely on the first approach.

Much more attention has been devoted to density waves than to bending waves in the assessment of these angular momentum exchanges between rings and satellites. This follows because the strongest torques for any given satellite are associated with density waves, from the remark made above that the strongest density waves (i.e., those with $k = 0$) are independent of the satellite's eccentricity and inclination), whereas the strongest bending wave amplitudes (for $k = 1$) are proportional to the satellite's inclination. Therefore, bending wave torques are expected to be negligible in the overall exchange of angular momentum between rings and any given satellite.

There does not appear to exist a direct evaluation of bending wave torques in the literature, from any of the communities dealing with disk-satellite interactions ("Satellite" here refers to any body orbiting the central object, be it a moon, a planet, a star, or a galaxy). However, both Bertin and Mark (1980) (in a galactic dynamics context) and Shu et al. (1983) (in a planetary ring context) provide expressions for the angular momentum luminosity, thereby quantifying the torques from bending waves according to the remark made above about the equality of the torque and the luminosity.

In these notes, we provide an alternative derivation of the torque. This derivation proceeds by noting that the generic torque expressions and linear solutions for density waves and bending waves are formally equivalent. Therefore, the density and bending wave torques are formally identical, *mutatis mutandis*. The exposition focuses on the leading order bending wave torques at inner vertical resonances (IVR) for the sake of definiteness, but this can easily be generalized to higher order torques and/or outer vertical resonances.

**A1. Inner Lindblad Resonances: a reminder**

Ring fluid particle positions in a density wave are given by (for details see Longaretti 1992 & 2017)

$$r = a\left[1 - e\cos\left(m(\varphi - \Omega_p t) + m\Delta_h\right)\right] \tag{A1}$$

where $\varphi$ is the fluid particle mean longitude, $e$ its eccentricity, $m\Delta_h$ the apsidal shift and $\Omega_p = \Omega_s + k\kappa_s/m$ the pattern speed associated with the satellite resonance ($\Omega_s$ and $\kappa_s$ are the epicyclic rotation velocity and epicyclic frequency of radial motions, respectively). An implicit choice of origins of time and angles has been made in this relation: at $t=0$, the satellite is at periapse and its mean epicyclic longitude is zero. As a consequence, for $k=0$ resonances, the origin of angles in the rotating frame is the satellite's mean epicyclic longitude.

The relevant component of the satellite potential giving rise to the considered resonance reads (see, e.g., Goldreich and Tremaine 1980, 1982; Shu 1984)

$$\phi(r,\theta,t) = \Phi_{m,k}(\beta)\cos m(\theta - \Omega_p t), \tag{A2}$$

where $\beta = a/a_s$ and $\Omega_p$ is the pattern speed associated with the satellite resonance. The satellite's dynamical action is characterized by (setting $r = a$ in $\Phi_{m,k}$)

$$\psi_{m,k} \equiv a\frac{d\Phi_{m,k}}{da} + 2m\Phi_{m,k} \tag{A3}$$

The strongest resonance corresponds to $k=0$. One has, for $m>1$ (Goldreich and Tremaine, 1980; Shu, 1984; Tiscareno & Harris 2017)

$$\Phi_{m,k=0} = -\frac{GM_s}{a_s}b_{1/2}^m(\beta), \tag{A4}$$

and

$$\frac{d\Phi_{m,k=0}}{da} = -\frac{GM_s}{a_s a}\beta Db_{1/2}^m(\beta), \tag{A5}$$

where $\beta=a/a_s$, $D \equiv d/d\beta$, and $b_\gamma^m(\beta)$ is a Laplace coefficient that is defined along with its derivatives (Murray & Dermott 1999, Eq. 6.67).

For a second order resonance ($k=1$) we have

$$\Phi_{m,k=1} = -\frac{GM_s}{a_s} e_s \left[\frac{1}{2} + m + \frac{1}{2}\beta D\right] b_{1/2}^m(\beta), \tag{A9}$$

and

$$\frac{d\Phi_{m,k=1}}{da} = -\frac{GM_s}{a_s a} e_s \left[(1+m)\beta D + \frac{1}{2}\beta^2 D^2\right] b_{1/2}^m(\beta), \tag{A10}$$

where $D^2 \equiv d^2/d\beta^2$.

The angular momentum per unit mass of ring fluid particles is

$$H = r^2 \frac{d\theta}{dt} = na^2 \left[1 - \frac{1}{2}\left(\frac{\kappa}{n}\right)^2 e^2\right] + O(e^4), \tag{A11}$$

and

$$\frac{dH}{dt} = rS_s, \tag{A12}$$

Here $S_s$ is the tangential component of the acceleration due to the satellite.

The torque $T$ follows from (Borderies et al. 1982; Shu et al. 1985; Longaretti 1992, 2017)

$$T = \int da\, d\varphi\, a\Sigma \frac{dH}{dt} = \int da\, \pi ma \psi_{m,k}\, \text{Im}(Z), \tag{A13}$$

where $Z = e\exp(im\Delta)$; this relation is exact (for both linear and nonlinear density waves). For a linear density wave at an ILR, $Z$ is given by (Shu 1984; Shu et al. 1985; Longaretti 1992)

$$Z = -i\frac{\psi_{m,k}}{2\pi G\Sigma a_r}\exp(ix^2/\delta)\int_{-\infty}^{x}\exp(-iy^2/\delta)dy \tag{A14}$$

where $x = (a - a_r)/a_r$ ($a_r$ is the resonance radius implicitly defined by Eq. 1). For $m > 1$

$$\delta = \frac{2\pi\Sigma a_r^2}{3(m-1)M_p}. \tag{A15}$$

Inserting Eq. (A4) in Eq. (A3) finally yields (Shu et al. 1985, Longaretti 1992)

$$T = -\frac{\pi^2 ma^3 (\psi_{m,k})^2 \Sigma}{3(m-1)GM_p}. \tag{A16}$$

The torque at an OLR has the same expression but an opposite sign. We recover the expression of the torque for a Lindblad resonance in Eq. (4) once the definition of the dimensionless quantity $A_{m,k}$ is introduced

$$A_{m,k} = -\frac{a_s \psi_{m,k}}{2GM_s}. \tag{A17}$$

In particular:

$$A_{m,k=0} = \frac{1}{2}[2m + \beta D]b_{1/2}^m, \tag{A18}$$

$$A_{m,k=1} = \frac{1}{2}\left[m(m+1) + (2+m)\beta D + \frac{1}{2}\beta^2 D^2\right]b_{1/2}^m. \tag{A19}$$

### A2. Inner Vertical Resonances

The position of any fluid particle in a bending wave can be described by

$$\begin{aligned} r &= a, \\ \theta &= \varphi, \\ z &= h = aI\sin[m(\varphi - \Omega_p t) + m\Delta] \end{aligned} \tag{A20}$$

where $\varphi$ is the fluid particle mean longitude. The azimuthal wavenumber $m$ is positive by assumption. The pattern speed $\Omega_p$ is discussed below. The resonance radius $a_r$ is implicitly defined by

$$m[n(a_r) - \Omega_p] = \mu(a_r), \tag{A21}$$

where $\mu = n - \dot{\Omega}$ is the frequency of vertical oscillations.

An implicit choice of the origins of time and angle has been made in the previous relations: $t = 0$ is the time of passage of the satellite through the equatorial plane and the origin of angles is the angular position of the satellite at $t = 0$. For simplicity, the satellite is assumed to orbit on a circular inclined orbit with inclination $i_s$. The theory of planetary motion adopted here is epicyclic, so that this orbit is precessing at a rate $\dot{\Omega}_s = n_s - \mu_s$ (see Borderies-Rappaport and Longaretti 1994 for a description of the epicyclic kinematics and dynamics in inclined orbits).

The form of the satellite potential for this type of resonance is less well known than that for a Lindblad resonance. Because deviations from the equatorial plane are small, one can write

$$\varphi_s(r,\theta,z,t) = \Phi_{s0}(r,\theta,z=0,t) + \left(\frac{\partial \phi_s}{\partial z}\right)_{z=0} z. \tag{A22}$$

Now, the vertical derivative of the potential is given by, for any term in the Fourier expansion (see Appendix A of Shu 1984 or section IIe of Shu et al. 1983)

$$\left(\frac{\partial \phi_s}{\partial z}\right)_{z=0} = \psi_{m,k}(r)\sin\left[m(\theta - \Omega_p t)\right]. \tag{A23}$$

The strongest vertical resonances correspond to $k = \pm 1$ (for $k = 0$, there is no vertical forcing, as the inclination of the satellite is neglected in this term). Inner vertical resonances (IVR) correspond to $\Omega_p = n_s + k\mu_s/m = \left[(m+k)n_s - k\dot{\Omega}_s\right]/m$ with $k = 1$ for the strongest ones ($k = -1$ for an OVR). For $k = \pm 1$ ($m > 1$ assumed),

$$\psi_{m,k}(r) = \frac{GM_s}{2a_s} b_{3/2}^m(\beta) I_s. \tag{A24}$$

The angular momentum per unit mass of ring fluid particles is

$$H = r^2 \frac{d\theta}{dt} = na^2\left[1 - \frac{1}{2}\left(\frac{\kappa}{n}\right)^2 I^2\right] + O(I^4). \tag{A25}$$

We have again

$$\frac{dH}{dt} = rS_s. \tag{A26}$$

The torque $T$ follows from

$$T = \int da \, d\phi \, a \Sigma \frac{dH}{dt} = \int da \, \pi ma \psi_{m,k} \, \text{Im}(Y), \tag{A27}$$

where $Y = I \exp(im\Delta)$ and Eqs. A10, A12 and A13 have been used to derive the second equality. This relation is also exact; note the analogy with Eq. A3. For a linear bending wave at an IVR, $Y$ is given by (Shu et al. 1983; Shu 1984)

$$Y = -i \frac{\psi_{m,k}}{2\pi G \Sigma a_r} \exp(ix^2/\delta) \int_{-\infty}^{x} \exp(-iy^2/\delta) dy, \tag{A28}$$

where $x = -(a - a_r)/a_r$ ($a_r$ is the resonance radius implicitly defined by Eq. A11) and $\delta$ is again defined by Eq. A5.

Note the formal equivalence between $Y$ in Eq. A17 and $Z$ in Eq. A4, except for the sign of $x$ (this change of sign reflects the fact that an ILR density wave and an IVR bending wave propagate in opposite directions). Because of this analogy, the torque is again given by

$$T = -\frac{\pi^2 ma^3 (\psi_{m,k})^2 \Sigma}{3(m-1)GM_p}. \tag{A29}$$

Here again, an OVR torque has the same magnitude but opposite sign.

**Acknowledgements**

We gratefully thank Peter Goldreich for a brief but valuable discussion on this topic. We also thank Joe Spitale for reviewing this paper, and M. Hedman and M. Tiscareno for fruitful discussions. RT and MEM were funded by the Cassini mission. PYL acknowledges support by the French National Program of Planetology (PNP).
**References:**

Bertin G. and Mark J.W.K., On the excitation of WARPS in galaxy disks, *Astron. Astrophys.*, **88** (3), 289-297 (1980).

Borderies N., Goldreich P. and Tremaine S., Sharp edges of planetary rings, Nature, 299, 209-211 (1982).

Borderies N., Goldreich P. and Tremaine S., Unsolved problems in planetary ring dynamics, *in: Planetary rings* (1984).

Borderies N., Goldreich P. and Tremaine S., A granular flow model for dense planetary rings, *Icarus*, **63**, 406-420 (1985).

Borderies N., Goldreich P. and Tremaine S., The formation of sharp edges in planetary rings by nearby satellites, *Icarus*, **80**, 344-360 (1989).

Borderies-Rappaport N. and Longaretti P.Y., Test particle motion around an oblate planet, *Icarus*, **107**, 129 (1994).

Brouwer D. and Clemence G.M., *Methods of celestial mechanics*. Academic Press, New York (1961).

Caudal, G. 2017, Icarus, 286, 280

Cooper N.J., Murray C.D., Lainey V. et al., Cassini ISS mutual event astrometry of the mid-sized Saturnian satellites 2005-2012, *Astron. Astrophys.*, **572**, A43 (2014).

Cooper N.J., Renner S., Murray C.D. and Evans M.W., Saturn's Inner Satellites: Orbits, Masses, and the Chaotic Motion of Atlas from New Cassini Imaging Observations, *Astron. J.*, **149** (1), id. 27 (2015).

Chakrabarti, S.K., The dynamics of particles in the bending waves of planetary rings. Mon. Not. R. Astron. Soc. 238, 1381–1394 (1989).

Charnoz S., Salmon J. and Crida A., The recent formation of Saturn's moonlets from viscous spreading of the main rings, Nature, 465 (7299), 752-754 (2010).

Charnoz S., Crida A., Castillo-Rogez J.C. et al. Accretion of Saturn's mid-sized moons during the viscous spreading of young massive rings: Solving the paradox of silicate-poor rings versus silicate-rich moons, *Icarus*, **216** (2), 535-550 (2011).

Crida A., Morbidelli A. and Masset F., On the width and shape of gaps in protoplanetary disks, *Icarus*, **181** (2), 587-604 (2006).

Dewar R.L., A Lagrangian Derivation of the Action-Conservation Theorem for Density Waves, *Astrophys. J.*, **174**, 301 (1972).

Daisaka H., Tanaka H. and Ida S., Viscosity in a Dense Planetary Ring with Self-Gravitating Particles, *Icarus*, **154** (2), 296-312 (2001).

Daisaka, H., & Ida, S. 1999, Earth, Planets and Space, 51, 1195

Dermott, S. F., Malhotra, R., & Murray, C. D. 1988, Icar, 76, 295

Durisen, R. H., Carmer, N. L., Murphy, B. W., et al. 1989, Icar, 80, 136

El Moutamid M., Nicholson P.D., French R.G. et al., How Janus' orbital swap affects the edge of Saturn's A ring? *Icarus*, **279**, 125-140 (2016).

El Moutamid M., Hedman M.M., Tajeddine R. et al., Using Saturn's Rings as a Diagnostic of its Internal Differential Rotation, *In Preparation* (2017).

Esposito, L.W., O'Callaghan, M., West, R.A., The structure of Saturn's rings: Implications from the Voyager stellar occultation. *Icarus* **56**, 439–452 (1983).

Estrada, P. R., Cuzzi, J. N., 1996, 122 (2), 251

Fuller, J., Luan, J., & Quataert, E. 2016, MNRAS, 458 (4), 3867

Goldreich P. and Tremaine S.D., The formation of the Cassini division in Saturn's rings, *Icarus*, **34**, 240-253 (1978a).

Goldreich P. and Tremaine S.D., The excitation and evolution of density waves, *Astrophys. J.*, **222**, 850-858 (1978b).

Goldreich P. and Tremaine S.D., The excitation of density waves at the Lindblad and corotation resonances by an external potential, *Astrophys. J.*, **233**, 857-871 (1979).

Goldreich P. and Tremaine S.D., Disk-satellite interactions, *Astrophys. J.*, **241**, 425-441 (1980).

Goldreich P. and Tremaine S.D., The dynamics of planetary rings, *Ann. Rev. Astron. Astrophys.*, **20**, 249-283 (1982).

Goldreich P., Rappaport N., Chaotic motions of Prometheus and Pandora, *Icarus*, **162** (2), 391-399 (2003).

French R.G., Elliot J.L., Allen D.A., Inclinations of the Uranian rings, Nature, 298, 827-829 (1982).

Hahn J.M., Spitale J.N. and Porco C.C., Dynamics of the Sharp Edges of Broad Planetary Rings, *Astrophys. J.*, **699** (1), 686-710 (2009).

Hedman M.M. and Nicholson P.D., Kronoseismology: Using Density Waves in Saturn's C Ring to Probe the Planet's Interior, *Astron. J.*, **146** (1), id.12 (2013).

Hedman M.M. and Nicholson P.D., More Kronoseismology with Saturn's rings, Mon. Not. R. Astron. Soc., 444 (2), 1369-1388 (2014).

Hedman M.M. and Nicholson P.D., The B-ring's surface mass density from hidden density waves: Less than meets the eye?, *Icarus*, **279**, 109-124 (2016).

Jacobson R.A., Antreasian P.G., Bordi J.J. et al., The Gravity Field of the Saturnian System from Satellite Observations and Spacecraft Tracking Data, *Astron. J.*, **132** (6), 2520-2526 (2006).

Jacobson R.A., Spitale J., Porco C.C. et al., Revised Orbits of Saturn's Small Inner Satellites, *Astron. J.*, **135** (1), 261-263 (2008).

Kaula W.M, Tidal Dissipation by Solid Friction and the Resulting Orbital Evolution, *Rev. Geophys. Space Phys.*, **2**, 661-685 (1964).

Lainey V., Karatekin Ö, Desmars J. et al., Strong Tidal Dissipation in Saturn and Constraints on Enceladus' Thermal State from Astrometry, *Astrophys. J.*, **752** (1), id.14 (2012).

Lainey V., Jacobson R.A., Tajeddine R. et al. New constraints on Saturn's interior from Cassini astrometric data, *Icarus*, **281**, 286-296 (2017).

Lesur, G and Longaretti P.Y., On the relevance of subcritical hydrodynamic turbulence to accretion disk transport. *Astron. Atrophys.*, **444,** 25-44 (2005).

Lin D.N.C. and Papaloizou J., Tidal torques on accretion discs in binary systems with extreme mass ratios, Monthly Notices of the Royal Astronomical Society, 186, 799-812 (1979).

Lissauer, J. J., Shu, F. H., and Cuzzi, J. N., Viscosity in Saturn's rings. In: Brahic, A. (Ed.), Planetary Rings, Proceedings of IAU Symposium No. 75, Toulouse, France, pp. 385–392 (1984).

Lissauer J.J., Bending waves and the structure of Saturn's rings, *Icarus*, **62**, 433-447 (1985).

Lissauer J.J., Goldreich P. and Tremaine S., Evolution of the Janus-Epimetheus coorbital resonance due to torques from Saturn's rings, *Icarus*, **64**, 425-434 (1985).

Longaretti P.Y., Planetary ring dynamics: from Boltzmann's equation to celestial mechanics, *In: Interrelations between physics and Dynamics for minor bodies in the solar system*, editors: Benest D. and Froeschle. Available on arXiv: astro-ph/1606.00759 (1992).

Longaretti P.Y., On the Phenomenology of Hydrodynamic Shear Turbulence, *Astrophys. J.*, **576** (1), 587-598 (2002).

Longaretti P.Y., Theory of Narrow Rings and Sharp Edges, In: *Planetary Ring Systems,* C.D Murray and M.S. Tiscareno, eds. Cambridge University Press. (2017a).

Longaretti P.Y., The dynamics of Saturn's B ring edge revisited 1. Forced edge mode and edge torque, *In preparation* (2017b).

Lynden-Bell D. and Kalnajs A.J., On the generating mechanism of spiral structure, *Mon. Not. R. Astron. Soc.*, **157**, 1 (1972).

Lynden-Bell D. and Pringle J.E., The evolution of viscous discs and the origin of the nebular variables, *Mon. Not. R. Astro. Soc.*, **168**, 603-637 (1974).

Marley M.S. and Porco C.C., Planetary acoustic mode seismology - Saturn's rings, *Icarus*, **106**, 508 (1993).

Meyer-Vernet N. and Sicardy B. On the physics of resonant disk-satellite interaction, *Icarus*, **69**, 157-175 (1987).

Murray C.D. and Dermott S.F. *Solar system dynamics*, Cambridge University Press, Cambridge, UK (1999).

Nicholson P.D. and Porco C.C., A new constraint on Saturn's zonal gravity harmonics from Voyager observations of an eccentric ringlet, *J. Geophys. Res.*, **93**, 10209-10224 (1988).

Nicholson P.D., French R.G., Hedman M.M., Marouf E.A. and Colwell J.E., Noncircular features in Saturn's rings I: The edge of the B ring, *Icarus*, **227**, 152-175 (2014).

Nicholson P.D., French R.G, and Spitale J.N., Narrow Rings, Gaps and Sharp Edges, In: *Planetary Ring Systems,* C.D Murray and M.S. Tiscareno, eds. Cambridge University Press. (2017).

Noyelles, B., Baillié, K., Lainey, V., & Charnoz, S. 2016, AAS DPS meeting #48, id.121.07

Noyelles, B. 2017, *Icarus*, 282, 276

Ogilvie, G. I., & Lin, D. N. C. 2004, ApJ, 610 (1), 477

Porco C.C., Danielson G.E., Goldreich P., Holberg J.B. and Lane A.L., Saturn's nonaxisymmetric ring edges at 1.95 R(s) and 2.27 R(s), *Icarus*, **60**, 17-28 (1984).

Porco C.C. and Goldreich P., Shepherding of the Uranian rings. I - Kinematics. II − Dynamics, *Astron. J.*, **93**, 724-737 (1987).

Prandt L. Bericht über Untersuchungen zur ausgebildeten Turbulenz, *Z. Angew. Math, Meth.*, **5**, 136-139 (1925).

Rhoden A.R., Henning W., Hurford T.A., Patthoff A.D. and Tajeddine R., The implications of tides on the Mimas ocean hypothesis. *J. Geophys. Res.*, **122** (2), 400-410 (2017).

Salmon J., Charnoz S., Crida A. and Brahic A., Long-term and large-scale viscous evolution of dense planetary rings. *Icarus*, **209** (2), 771-785 (2010).

Salo H., Schmidt J. and Spahn F., Viscous overstability in Saturn's B ring. 1. Direct simulations and measurement of transport coefficients, *Icarus,* **153**, 295-315 (2001).

Schmidt J., Salo H., Spahn F. and Petzschmann, O., Viscous overstability in Saturn's B ring. 2. Hydrodynamic theory and comparison to simulations, *Icarus,* **153**, 316-331 (2001).

Schmidt J., Ohtsuki K., Rappaport N., Salo H. and Spahn F., Dynamics of Saturn's Dense Rings, in: Saturn from Cassini-Huygens, 413 (2009).

Shu F.H., Cuzzi J.N. and Lissauer J.J. Bending waves in Saturn's rings, *in: Saturn Conference*, **53**, 185-206 (1983).

Shu F.H., Waves in planetary rings, *in: Planetary rings* (1984).

Shu, F.H., Dones, L., Lissauer, J.J., Yuan, C., Cuzzi, J.N., Non-linear spiral density waves: Viscous damping. Astrophys. J. 299, 542–573 (1985).

Spitale J.N. and Porco C.C., Time Variability in the Outer Edge of Saturn's A-Ring Revealed by Cassini Imaging, *Astron. J.*, **138** (5), 1520-1528 (2009).

Spitale J.N. and Porco C.C., Detection of Free Unstable Modes and Massive Bodies in Saturn's Outer B Ring, *Astron. J.*, **140** (6), 1747-1757 (2010).

Sremčević M., Schmidt J., Salo H. et al., A belt of moonlets in Saturn's A ring, Nature, 449 (7165), 1019-1021 (2007).

Tajeddine R., Cooper N.J., Lainey V., Charnoz S. and Murray C.D., Astrometric reduction of Cassini ISS images of the Saturnian satellites Mimas and Enceladus, *Astron. Astrophys.*, **551**, A129 (2013).

Tajeddine R., Lainey V., Cooper N.J. and Murray C.D., Cassini ISS astrometry of the Saturnian satellites: Tethys, Dione, Rhea, Iapetus, and Phoebe 2004-2012, *Astron. Astrophys.*, **575**, A73 (2015).

Tajeddine R., Nicholson P.D., Tiscareno M.S. et al., Dynamical phenomena at the inner edge of the Keeler gap, *Icarus*, **289**, 80-93 (2017).

Tajeddine, R. T., Rambaux, N., Lainey, V., et al. 2014, Science, 346 (6207), 322

Thomas, P. C. 2010, Icarus, 208 (1), 395

Tiscareno M.S., Burns J.A., Nicholson P.D., Hedman M.M. and Porco C.C., Cassini imaging of Saturn's rings II. A wavelet technique for analysis of density waves and other radial structure in the rings, *Icarus*, **189**, 14–34 (2007).

Tiscareno M.S., Burns J.A., Hedman M.M. et al., 100-metre-diameter moonlets in Saturn's A ring from observations of 'propeller' structures, *Nature*, **440** (7084), 648-650 (2006).

Tiscareno M.S., Burns J.A., Hedman M.M. and Porco C.C., The Population of Propellers in Saturn's A Ring, *Astron. J.*, **135** (3), 1083-1091 (2008).

Tiscareno M.S. and Harris B.E., Mapping spiral waves and other radial features in Saturn's rings. In preparation (2017).

Toomre A., Group Velocity of Spiral Waves in Galactic Disks, *Astrophys. J.*, **158**, 899 (1969).

Weiss J.W., Porco C.C. and Tiscareno M.S., Ring Edge Waves and the Masses of Nearby Satellites, *Astron. J.*, **138** (1), 272-286 (2009).

Yasui Y., Ohtsuki K. and Daisaka H. Viscosity in Planetary Rings with Spinning Self-gravitating Particles, *Astron. J.,* **143** (5), id. 110 (2012).